\documentclass{aa}  

\usepackage{graphicx}
\usepackage{txfonts}

\usepackage{natbib}
\bibpunct{(}{)}{;}{a}{}{,} 

\begin{document}

   \title{Coupling hydrodynamics with comoving frame radiative transfer}

   \subtitle{II. Stellar wind stratification in the high-mass X-ray binary Vela X-1}

   \author{A. A. C. Sander\inst{1}
        \and
           F. F{\"u}rst\inst{2}
        \and
           P. Kretschmar\inst{2}
        \and
           L. M. Oskinova\inst{1}
        \and
           H. Todt\inst{1}
        \and
           R. Hainich\inst{1}
        \and
           T. Shenar\inst{1}
        \and
           W.-R. Hamann\inst{1}
          }

   \institute{Institut f\"ur Physik und Astronomie, Universit\"at Potsdam,
              Karl-Liebknecht-Str. 24/25, D-14476 Potsdam, Germany\\
              \email{ansander@astro.physik.uni-potsdam.de}
              \and
              European Space Astronomy Centre (ESA/ESAC), Science Operations Department, Villanueva de la Cañada (Madrid), Spain
             }

   \date{Received July 17, 2017; accepted November 17, 2017}

\abstract{
 Vela X-1, a prototypical high-mass X-ray binary (HMXB), hosts a neutron star (NS) in 
 a close orbit around an early-B supergiant donor star. Accretion of the donor star's wind onto the NS powers its strong X-ray luminosity. To understand the physics of HMXBs, detailed knowledge about the donor star winds is required.  
}{
  To gain a realistic picture of the donor star in Vela X-1, we constructed a  hydrodynamically consistent atmosphere model describing the wind stratification  while properly reproducing the observed donor spectrum. To investigate how X-ray illumination affects the stellar wind, we calculated additional models for different X-ray luminosity regimes.
}{
  We used the recently updated version of the PoWR code to consistently solve the hydrodynamic equation together with the statistical equations and the radiative transfer. 
}{
The wind flow in Vela X-1 is driven by ions from various elements, with \ion{Fe}{iii} and \ion{S}{iii} leading in the outer wind. The model-predicted  mass-loss rate is in line with earlier empirical studies. The mass-loss rate is almost unaffected by the presence of the accreting NS in the wind. The terminal wind velocity is confirmed at $\varv_\infty \approx 600\,$\,km\,s$^{-1}$. On the other hand, the wind velocity in the inner region where the NS is located is only $\approx 100\,$\,km\,s$^{-1}$, which is not expected on the basis of a standard $\beta$-velocity law. In models with an enhanced level of X-rays, the velocity field in the outer wind can be altered. If the X-ray flux is too high, the acceleration breaks down because the ionization increases.
}{
  Accounting for radiation hydrodynamics, our Vela X-1 donor atmosphere model reveals a low wind speed at the NS location, and it provides quantitative information on wind driving in this important HMXB.   
}


\maketitle


\section{Introduction}
  \label{sec:intro}

  High-mass X-ray binaries (HMXBs) consist of a compact object -- either a neutron star (NS) or a black hole -- 
  that accretes material from a massive donor star. They are therefore a unique link between different important
  astrophysical fields, combining high-energy astrophysics and accretion with stellar outflows and winds.
  An especially interesting subclass of HMXBs are the so-called ``wind-fed'' systems where the compact object 
  accretes material directly from the stellar wind of the donor \citep[see][for a recent review on this subclass]{HMXBReview2017}.
  The prototype of such systems is Vela X-1, discovered by \citet{Chodil+1967}, with 
  its B-supergiant donor \object{HD 77581}. To avoid confusion, we hereafter refer to the system as \object{Vela X-1}, stating explicitly whether 
  we refer to the donor star or the NS when necessary. The adopted   parameters for the Vela X-1 system used or discussed 
  throughout this work are compiled in Table\,\ref{tab:sysparams}.
    
  Vela X-1  is a persistent X-ray source with a  typical luminosity a few times  $10^{36}$~erg~s$^{-1}$. 
  The X-ray source displays significant variability, including bright flares and very low, or ``off'', states \citep[e.g.,][]{Kreykenbohm+2008,Martinez-Nunez+2014}.
  Vela X-1 is an eclipsing binary, which provides the rare opportunity of studying the wind of the donor star during the X-ray eclipse.  
  Observations of the system during the eclipse with various instruments \citep{Sato+1986,Nagase+1994,Sako+1999,Schulz+2002} provide evidence of 
  optically thick and clumped matter in addition to warm ionized plasma.
  The mean flux and variability are explained by accretion from a wind with a complex structure, including clumps, 
  turbulent motion, and larger structures \citep[e.g.,][]{Fuerst+2010,Manousakis+Walter2015}. 
  A precise knowledge about the donor star and its wind parameters is 
  essential for studying these hypotheses in detail and for understanding wind-fed HMXBs in general.
    
  The X-ray variability of Vela X-1 has recently been modeled by \citet{Manousakis+Walter2015}  using 
  the 2D hydrodynamics code VH-1 \citep{Blondin+1990,Blondin+1991,Blondin+Pope2009}.  These elaborate multidimensional hydrodynamics codes allow for complex geometries, but they 
  treat the donor wind in an approximate way, for instance, by using
  the Sobolev approximation of a CAK radiative force \citep{Castor+1975, Blondin+1990} and an ionization parameter $\xi$.  On the other hand, 
  sophisticated stellar atmosphere models allow for a detailed study of the line-driven donor wind, accounting for a variety of elements 
  with a multitude of levels and a detailed radiative transfer without assuming a local thermodynamical equilibrium (non-LTE). However, 
  such sophisticated model stellar atmospheres are restricted to a one-dimensional description. Therefore, both approaches, multi-D hydrodynamic 
  models and sophisticated stellar atmosphere models are truly complimentary. 
                
  The donor wind of Vela X-1 was recently analyzed by \citet{GimenezGarcia+2016}, using for the first time detailed expanding stellar atmosphere 
  models for radiation-driven winds. Their results provided important indications of a potential dichotomy  between wind properties in classical
  persistent supergiant X-ray binaries (SGXBs) and  the so-called supergiant fast X-ray transients (SFXTs), which exhibit a significant variation 
  in their X-ray luminosity between quiescence and outbursts. 
        
\begin{table}
  \caption{Selected Vela X-1 system parameters}
  \label{tab:sysparams}
  \centering
  \begin{tabular}{l c c}
  \hline\hline
    Parameter                                                               &  Value                  &  Ref.  \\
     \hline
    orbital period $P_\text{orb}$\,[days]  \rule[0mm]{0mm}{3mm}             & $8.964357$              &  (1)   \\
    eccentricity $e$                                                        & $0.0898$                &  (2)   \\
    orbital separation $d_\textsc{ns}$\,[cm]                                & $3.5 \cdot 10^{12}$     &  (3)   \\
    rotational velocity $\varv_\text{rot} \cdot \sin i$\,[km\,s$^{-1}$]     & $56$                    &  (4)   \\
    orbital speed of the NS $\varv_\text{orb}$\,[km\,s$^{-1}$]              & $284$\tablefootmark{a}  &   --   \\
     \hline
  \end{tabular}
  \tablefoot{
     \tablefoottext{a}{calculated from $P_\text{orb}$ and $d_\textsc{ns}$, assuming a circular orbit}\\
     References. (1) \citealp{Kreykenbohm+2008}; (2) \citealp{Bildsten+1997}; (3) \citealp{GimenezGarcia+2016}; (4) \citealp{Fraser+2010}
  }  
\end{table} 
        
        The models used in \citet{GimenezGarcia+2016} assume a prescribed wind velocity field, and hence measured only 
        the terminal wind velocity $\varv_\infty$ precisely. However, more important in terms of accretion onto an NS is of 
        course the wind velocity at the location of the NS $d_\textsc{ns}$. For Vela X-1, \citet{vanKerkwijk+1995} determined 
        $d_\textsc{ns} \sim53\,R_\odot$ or $\sim1.8\,R_\ast$ at periastron; this is a relatively common value for such systems \citep[e.g.,][]{Falanga+2015}. 
    
        When the NS is only about one stellar radius away from the donor, the wind velocity at the distance of the NS $\varv(d_\textsc{ns})$ is
        much lower than its terminal value $\varv_\infty$ and strongly depends on the shape of the velocity field. Stellar atmosphere 
        models normally do not have a self-consistent wind stratification, but instead assume a stratification given by a so-called $\beta$-law, that is,
  \begin{equation}
    \label{eq:betasimple}
    \varv(r) = \varv_\infty \left( 1 - \frac{R_\ast}{r} \right)^{\beta}\text{.}
  \end{equation}
        While this is usually sufficient to measure the stellar and wind parameters quite accurately, it essentially means that 
        in these models the balance between inward- and outward-pushing forces is usually violated, and the assumed 
        velocity field would not be obtained when solving the hydrodynamic equation of motion. 
        In many cases, this level of consistency is not necessary. However, as soon as not only the global stellar and 
				wind parameters are of interest, but the particular physical properties throughout the
        stratification, especially closer to the star, the use of such an approximate treatment can lead to significant errors in the deduced
        properties. To overcome this problem, we present a hydrodynamically self-consistent atmosphere model for the donor of Vela X-1, using the 
        method recently presented in \citet{Sander+2017} for a new generation of models developed with the Potsdam Wolf-Rayet (PoWR) code. 
    
        A comparable approach has been used by \citet{Krticka+2012}, who calculated a set of 1D wind models for different orbit inclination angles. 
        This and their follow-up work focused on the angle-dependent behavior and a parameter-space study \citep{Krticka+2015}, while stellar parameters 
        were adopted from previous literature and no spectral cross-check of the results was performed. In this work, we focus on obtaining a hydrodynamically 
        self-consistent solution for the wind structure and on comparing our results with observed optical/UV spectra. The goal is to obtain a detailed wind 
        stratification tailored for the Vela X-1 donor. 

        This approach allows us to check whether the relatively low value of $\varv_\infty \approx 700\,$km\,s$^{-1}$ measured by \citet{GimenezGarcia+2016}
        can be explained by radiative driving alone, or if additional mechanisms have to be taken into account, such as\ the influence of X-ray irradiation 
        of the donor wind. Furthermore, we can qualitatively mimic the orbital modulation of the UV wind lines as originally predicted by \citet{Hatchett+1977}
        and compare the model with observations to gain further insight on the wind structure. In this paper we present hydrodynamically consistent stellar 
        atmosphere solutions for three test cases and also provide the resulting wind stratifications, especially for potential use in further studies.
                
        In Sect.\,\ref{sec:powr} we briefly summarize the physics applied in the PoWR models. 
        The following Sect.\,\ref{sec:results} then discusses the results of the modeling, with subsections focusing on
        the differences compared to the model from \citet{GimenezGarcia+2016}, using a prescribed mass-loss rate and velocity field,
        a study of the X-ray influence, and the discussion of the wind driving and the particular effect of the X-rays on it
        for the Vela X-1 donor. Finally, the conclusions are drawn in Sect.\,\ref{sec:conclusions}.

\section{PoWR}
  \label{sec:powr}

  \subsection{Fundamental concepts and parameters}
    \label{sec:basics}

        The Potsdam Wolf-Rayet (PoWR) model atmosphere code \citep[e.g.,][]{GKH2002,HG2003,Sander+2015} allows calculating stellar atmosphere models
        for a spherically symmetric star with a stationary mass outflow. The intricate non-LTE conditions in
        these atmospheres are properly accounted for by performing the radiative transfer in the comoving 
        frame (CMF) and obtaining the population numbers from the equations of statistical equilibrium. As these
        two are tightly coupled, they are iteratively updated together with the electron
        temperature stratification. The latter is required to ensure energy conservation in the expanding 
        atmosphere and can be obtained with the Uns{\"o}ld-Lucy method \citep{Lucy1964,Unsoeld1955} generalized for expanding atmospheres \citep{HG2003},
        or alternatively, through the electron thermal balance \citep[][]{Kubat+1999,Kubat2001}. 
        In the parameter regime used in this work, the electron thermal balance combined with the flux 
        consistency terms from the Uns{\"o}ld-Lucy method were found to be most effective in order to gain 
        a stable and reliable temperature stratification.

        Following the empirical solution for the Vela X-1 donor by \citet{GimenezGarcia+2016}, we defined
        our PoWR models for this work by the following basic input parameters: a stellar temperature
        ($T_\ast$) at a radius where the Rosseland continuum optical depth is $\tau_\text{Ross} = 20$, a 
        luminosity ($L$), and a stellar mass ($M_\ast$). The stellar radius $R_\ast$ is then defined thruogh 
        the Stefan-Boltzmann law ($L = 4\pi\sigma_\text{SB}T_\ast^4 R_\ast^2$), and the surface gravity 
        $g_\ast = g(R_\ast)$ immediately follows from $g_\ast = G M_\ast R_\ast^{-2}$. A full list of
        possible input parameter combinations for PoWR models is given in \citet{Sander+2017}. Depending
        on the literature and also on the model atmosphere code, the definitions for $R_\ast$ and the corresponding
        effective temperatures can vary. In the PoWR models, $R_\ast$ refers to the inner boundary of
        the model calculations and is therefore set so far inward that at this depth, a quasi-hydrostatic equilibrium
        is usually guaranteed. The effective temperature referring to $R_\ast$ is consequently noted as $T_\ast$.
        In addition to this modeling-motivated definition, the term ``effective temperature'' is also used for
        the temperature corresponding to the ``photospheric radius'', which is typically defined at the radius $R_{2/3}$ with an
        optical depth of $\tau_\text{Ross} = 2/3$. For OB stars, the difference between $R_\ast$ and $R_{2/3}$ is
        usually very small, and the resulting differences between $T_\ast$ and $T_{2/3}$ are on the order
        of about $1\,$kK for supergiants (see Sect.\,\ref{sec:results} for the values of $T_{2/3}$ for our models here). 
        However, these differences can be significantly larger for denser atmospheres, for example, in Wolf-Rayet
        stars, where a factor of two between $T_\ast$ and $T_{2/3}$ is common. 
        To avoid any confusion between the two reference systems, we refrain from using $T_\text{eff}$ in this work
        and always state either $T_\ast$ and $R_\ast$ or $T_{2/3}$ and $R_{2/3}$. The corresponding 
        surface gravity is denoted as $g_{2/3} = g(R_{2/3})$.
                
        Density inhomogeneities can be accounted for in the form of optically thin clumps with a void interclump
        medium and described by a so-called ``density contrast'' $D(r)$, sometimes also termed ``clumping factor'' $f_\text{cl}$, 
        assuming that the wind consists of clumps with a density $D(r)\cdot\rho(r)$ compared to a smooth wind with a density $\rho(r)$ \citep[][]{HK1998}.
        For a void interclump medium, its inverse $f_\text{V}(r) = D^{-1}(r)$ is known as the ``volume filling factor''.
        As suggested by the notation, the density contrast      can be depth dependent. Various depth-dependent parametrizations
        exist, and since their different effects
        easily merit their own dedicated paper, we stick to one parametrization for all hydrodyamically consistent 
        models used here. In our case, we used a similar approach as for the O supergiant model 
        in \citet{Sander+2017}, namely
        $D(r) = 1/f_\text{V}(r),$ with
  \begin{equation}
     \label{eq:clumpstrat}
      f_\text{V}(r) = f_{\text{V},\infty} + (1 -  f_{\text{V},\infty}) \cdot \exp\left(-\frac{\tau_\text{cl}}{\tau_\text{Ross}(r)}\right)\text{,}
  \end{equation}  
        that is, we assumed an essentially unclumped atmosphere at the inner boundary with a smooth transition to a significantly
        clumped wind in the outer part. The parameter $\tau_\text{cl}$, which is set to $0.1$ in this work unless
        otherwise noted, does not mark a strict transition zone, but instead describes
        a characteristic value for the (noticeable) onset of the clumping. The maximum value of $D_\infty = 11$
        is taken from \citet{GimenezGarcia+2016}. They did not use Eq.\,(\ref{eq:clumpstrat}), but a 
        different clumping parametrization with two velocities describing the region where the clumping factor increases.
        Unfortunately, a parametrization connecting density contrasts with explicit velocities is numerically unfavorable in the
        case of updates of $\varv(r)$, which is why we changed to Eq.\,(\ref{eq:clumpstrat}) for this work.
        
        For a converged atmosphere model, the synthetic spectrum is calculated using a formal integration in the observer's frame.
        The resulting spectrum is then convolved to account for the rotational broadening of the lines that is due to a projected rotation velocity 
        of $\varv_\text{rot} \sin i = 56\,$km\,s$^{-1}$ \citep{Fraser+2010}, as denoted in Table\,\ref{tab:sysparams}. This value is lower than
        earlier measurements from \citet{Zuiderwijk1995} and \citet{Howarth+1997} and also lower than calculations from \citet{Falanga+2015},
        but well motivated by the clearly unblended optical \ion{O}{ii} lines, as demonstrated in \citet{GimenezGarcia+2016}.
        Since this rotational speed is also very different from the critical velocity ($\varv_\text{rot} \sin i /\varv_\text{rot,crit} \approx 0.15$), the convolution
        method should be sufficient, and we therefore do not account for rotational effects during our atmosphere calculations.

 \subsection{Hydrodynamic branch}
          \label{sec:powrhydro}
        
        In contrast to standard PoWR models, where the mass-loss rate $\dot{M}$ and the velocity field $\varv(r)$
        are prescribed in order to essentially measure $\dot{M}$ and the terminal velocity $\varv_\infty$, hydrodynamically
        consistent models predict these values by including a depth-dependent solution of the hydrodynamic equation of motion
        in the main iteration scheme. In a particular hydrodynamic stratification update, the velocity field is obtained
        by integrating inward and outward from the critical point,  while the mass-loss rate itself is implicitly fixed 
        by requiring a velocity field that is continuous in $\varv(r)$ and $\mathrm{d}\varv/\mathrm{d}r$.

        Although $\dot{M}$ and $\varv(r)$ are output quantities, we need to assign them with initial values for
        the iteration. For this purpose, we used a non-hydrodynamic model that we created based on
        the results of \citet{GimenezGarcia+2016}. 

        In non-hydrodynamical (non-HD) models that are used for empirical studies, only those elements and ions usually need to be considered that
        either leave an imprint in the spectral appearance, or have a significant influence through blanketing. For hydrodynamically
        self-consistent models, however, all the elements and ions that contribute in a non-negligible way to the radiative
        acceleration have also to be taken into account. Together with the additional time required for convergence due to the stratification
        updates, this is a second factor that makes these models numerically more costly. With a total of 11 elements,
        the model for Vela X-1 used in \citet{GimenezGarcia+2016} is quite large to begin with. For our purpose in this work,
        Ne, Cl, Ar, K, and Ca were added in various ionization stages, which increased the total number of considered elements 
        to 16. A detailed list of the ions and lines we used in the radiative transfer calculations
        is provided in Table\,\ref{tab:datom}. For a model with a larger X-ray component, we also need to account for more of
        the higher ionization stages. In order to keep the overall number of levels manageable, we reduced the number of levels 
        in several of the lower ionization stages before adding the higher ions. We have cross-checked with test calculations that the fewer 
        levels in the lower stages do not notably affect the obtained radiative acceleration $a_\text{rad}$ as long as the model
        parameters are the same. This alternative set of atomic data, including the higher ions, is listed in parentheses in Table\,\ref{tab:datom}. 
        
        The approach for obtaining HD consistent PoWR models of the current generation is extensively 
        described in \citet{Sander+2017}, including an example application to an O4 supergiant ($\zeta\,$Pup/\object{HD$\,66811$}). 
        The application to the donor star of Vela X-1 now shows that the method also works in 
        the much cooler wind regime of an early B-type star.

 \subsection{Inclusion of X-rays}
          \label{sec:powrxray}
        
        The PoWR models can currently account for the effect of X-rays by assuming a hot and optically thin plasma 
        embedded in the cool wind. Instead of adding an additional X-ray component to the
        radiation field that is obtained by the radiative transfer, such as has been done in the models by \citet{Krticka+2012}, we 
        considered them as additional emissivities for each level $i$ at each frequency $\nu$,
        \begin{equation}
          \label{eq:etax}
          \eta_{\text{X},i,\nu} = X_\text{fill}~n_\text{e,X}~n_i~\sigma_{\text{ff},i}(T_\text{X},\nu)~\exp\left(-\frac{h\nu}{k_\text{B}T_\text{X}}\right)
        ,\end{equation}
        which were added before performing the radiative transfer calculations. The method, which was first introduced in \citet{Baum+1992}, 
        requires three parameters to be specified: the temperature $T_\text{X}$ of the hot component, the fraction with regard to
        the cool wind component $X_\text{fill}$, and the onset radius $R_0$. The parameters adapted in this work are compiled in 
        Table\,\ref{tab:modelparam} together with the results for each model. As indicated by the notation of the 
        cross section $\sigma_\text{ff}$ in Eq.\,(\ref{eq:etax}), only
        free-free emission (bremsstrahlung) is considered, with
        \begin{equation}
          \label{eq:sigmaff}
          \sigma_{\text{ff},i} = C_\text{ff}~Z_i^2~\nu^{-3}~T_\text{X}^{1/2}
  \end{equation}
        denoting $Z_i$ as the charge of the ion corresponding to level $i$, and $C_\text{ff}$ as the coefficient for the 
        free-free cross section \citep[see, e.g.,][]{Allen1973}. These additional X-ray emissivities were only added for 
        $r > R_0$. By adding the X-rays on the level of the emissivities, their influence is implicitly considered in all quantities 
        that are obtained during the radiative transfer calculations. Thereby, the X-rays automatically affect the resulting radiative rates and
        thus the population numbers, most notably by Auger ionization. 
        
\begin{figure}[thb]
  \resizebox{\hsize}{!}{\includegraphics[angle=270]{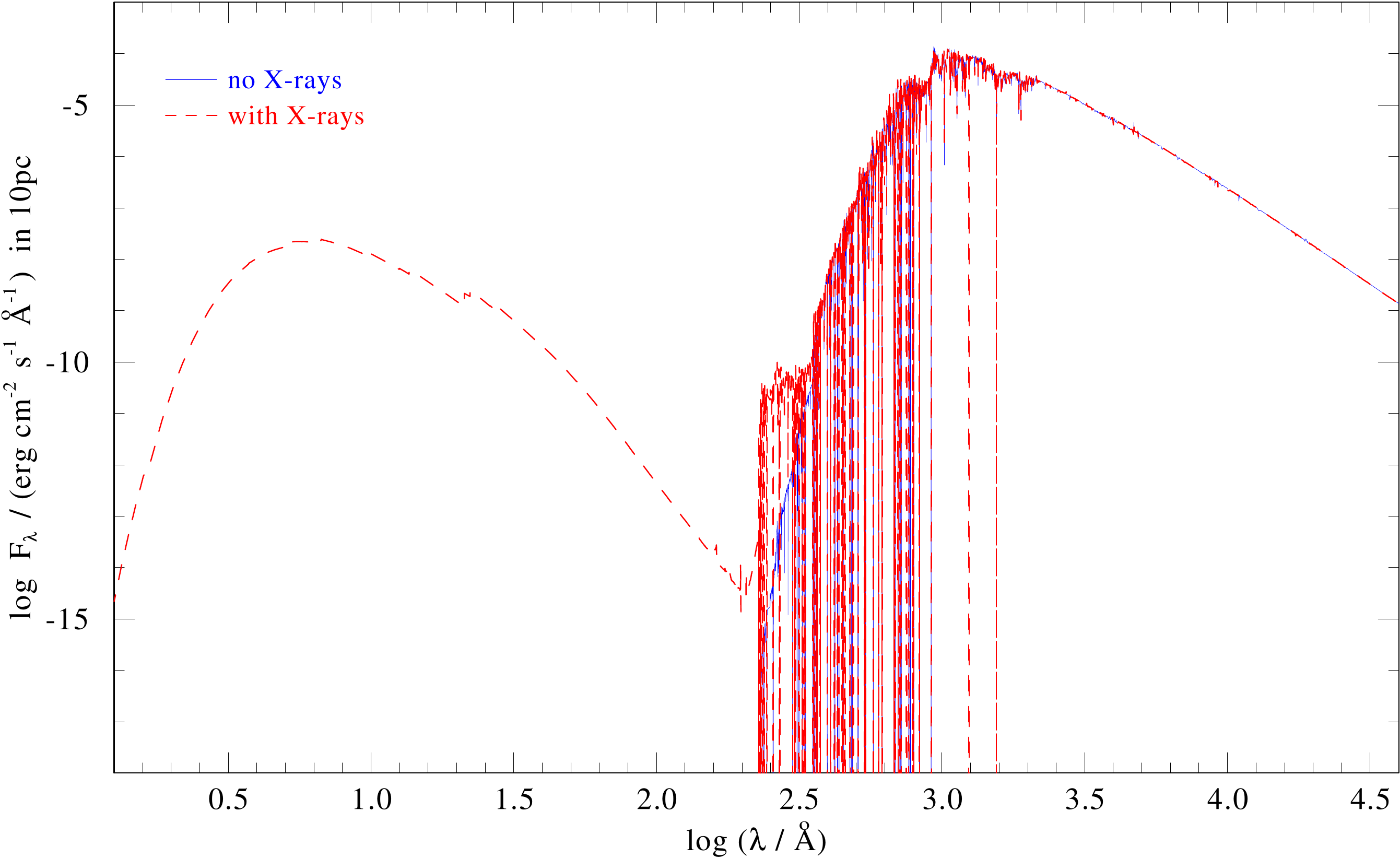}}
  \caption{Spectral energy distribution for a model without X-rays (blue curve) overplotted with those of
                 a model including X-rays using the moderate illumination parameters (red dashed curve).}
  \label{fig:xraysed}
\end{figure}

        An example of the emerging flux including our X-rays is shown in Fig.\,\ref{fig:xraysed}, where we
        compare the spectral energy distribution of a model with and without X-rays. The flux below the \ion{He}{ii}
        ionization edge drops to virtually zero in the model without X-rays. However, the effect of including them is not limited
        to this region, but also affects longer wavelength, most notably the regime directly redward of the edge.
        The normalized spectra of our Vela X-1 models also clearly
        show the effect of X-rays, for instance, in the \ion{N}{v} (1238\,\AA, 1242\,\AA) doublet, which is much 
				stronger than in a corresponding
        model that does not consider X-rays (see Figure~\ref{fig:uvlines}). 

        The shortcoming of this approach is that the hot  plasma is assumed to be distributed throughout the wind. However, 
				the X-rays resulting from the wind accretion onto an NS are generated only locally, thereby breaking the spherical symmetry. 
				Unfortunately, the sophisticated nature of modeling expanding stellar atmospheres so far prevents the detailed use and general application of 
        multidimensional approaches. First efforts in this field have been made \citep[see, e.g.,][]{HB2006,HB2014}, but
        the computational effort is overwhelmingly high for a general application, even when using significantly fewer 
        elements and levels than required for our present task. Thus, performing a detailed radiative transfer is currently limited to 
        1D model atmospheres, knowing that our consideration of the NS effects can only be a rather simple approximation.
        
        In this work, we used various options to include X-rays. After obtaining an initial HD-consistent stratification for a model 
        without X-rays, we first used the same ad hoc parameters as were assumed by \citet{GimenezGarcia+2016}, which were motivated 
        chiefly by the aim to reproduce the observed spectrum. 
        Since the X-ray luminosity they needed to assume was much lower than the average observed X-ray luminosity ($L_\text{X}$) of 
        Vela X-1, it was interpreted as intrinsic wind X-ray emission (which might stem from\ wind shocks, e.g.). However, the typical 
        X-ray luminosity of OB-stars is nearly two orders of magnitude lower than the 
        $L_\text{X} \approx 5.5 \cdot 10^{33}\,$erg\,s$^{-1}$ ($\log L_\text{X}/L_\text{bol} \approx -5.3$) assumed 
        by \citet{GimenezGarcia+2016}. The intrinsic wind X-ray emission is best assessed from the observations during the eclipse, for 
        which \citet{Schulz+2002} derived $L_\text{X} \approx 2.2 \cdot 10^{33}\,$erg\,s$^{-1}$. 
        We resolved to use this value for potentially intrinsic wind emission, although this is probably an overestimation because
        X-rays from the photoionized region around the accretion source can still be in the line of sight even during eclipse as this 
        region could exceed the area shielded by the donor star \citep[e.g.,][]{Watanabe+2006}.
    
        Finally, we calculated models where in addition to the intrinsic wind X-ray emission we used a second and significantly 
        larger X-ray component that in an approximate way describes the direct X-ray emission from the accreting NS. The observed spectrum 
        of the direct component is typically well described as a power law. We approximated the observed X-ray spectrum as a bremsstrahlung, 
        described by a suitable temperature $T_{\rm X}$, and used it in our modeling.  
        The resulting total $L_\text{X}$, that is, the luminosity based on the integrated emergent flux up to $124\,$\AA,  
        corresponds to a typical value for Vela X-1 that is fully sufficient for our study. 
        Because of the caveats in our modeling of the geometry of the X-ray emitting region around the NS,  
        our study is rather qualitative, and the detailed numbers should be taken with care.

\section{Results}
  \label{sec:results}

        To a obtain  starting model, we a calculated a non-HD model with the extended atomic data, but without any X-rays.
        Adopting the same notation as in \citet{Sander+2017}, where $P(r)$ is the sum of the gas and turbulence pressure, $\rho(r)$ is the density, and $\varv(r)$ the wind
        velocity, we obtained a \emph{\textup{work ratio}}
  \begin{equation}
    Q := \frac{\dot{M} \int \left(  a_\text{rad} - \frac{1}{\rho} \frac{\mathrm{d} P}{\mathrm{d} r} \right) \mathrm{d} r}{\dot{M} \int \left( \varv \frac{\mathrm{d} \varv}{\mathrm{d} r} + \frac{G M_{\ast}}{r^2} \right) \mathrm{d} r }\text{,}
  \end{equation}
        which is a measure for the integrated HD balance, of $Q \approx 1.05$. As described in the previous
        section, the clumping stratification was also changed compared to \citet{GimenezGarcia+2016}, so that our model slightly 
        differs from the empirical study in some parameters. The full list of input parameters we used in all HD models
        is given in Table\,\ref{tab:modelinput}. The models also differ in various parameters from those used in the study by \citet{Krticka+2012},
        who derived a mass-loss rate using a comparable approach, but accounted in their METUJE code only for the pure CMF line force \citep[see][]{KK2010}. 
        Since they did not provide any emerging spectrum, we preferred the stellar parameters from the empirical results by \citet{GimenezGarcia+2016}.

\begin{table}
  \caption{Input parameters for the Vela X-1 donor models}
  \label{tab:modelinput}
  \centering
  \begin{tabular}{l c c c}
  \hline\hline
     Parameter  \rule[0mm]{0mm}{3mm}                      &  \multicolumn{3}{c}{Value}     \\
        \hline
    $T_*$\,[kK]  \rule[0mm]{0mm}{3mm}                     & \multicolumn{3}{c}{  $25.5$ }  \\
    $R_*$\,[$R_\odot$]                                    & \multicolumn{3}{c}{  $28.4$ }  \\
    $\log\,(L$\,[$L_\odot$]$)$                             & \multicolumn{3}{c}{ $5.485$ }  \\
    $M_\ast$\,[$M_\odot$]                                 & \multicolumn{3}{c}{  $20.2$ }  \\
    $\log\,(g$\,[cm\,s$^{-2}$]$)$                          & \multicolumn{3}{c}{  $2.84$ }  \\
    $D_\infty$                                            & \multicolumn{3}{c}{    $11$ }  \\
    \medskip
    $\varv_\text{mic}$\,[km\,s$^{-1}$]                            & \multicolumn{3}{c}{    $10$ }  \\                 
    \textit{element}                                      & \multicolumn{2}{c}{\textit{mass fraction}} & \textit{rel.~ab.\tablefootmark{a}}       \\
    $X_\text{H}$\tablefootmark{b} \rule[0mm]{0mm}{3mm}    & \multicolumn{2}{c}{$0.65$}                 &  $0.89$  \\
    $X_\text{He}$\tablefootmark{b}                        & \multicolumn{2}{c}{$0.336$}                &  $1.35$  \\
    $X_\text{C}$\tablefootmark{b}                         & \multicolumn{2}{c}{$5.0 \times 10^{-4}$}   &  $0.17$  \\
    $X_\text{N}$\tablefootmark{b}                         & \multicolumn{2}{c}{$1.8 \times 10^{-3}$}   &  $2.11$  \\
    $X_\text{O}$\tablefootmark{b}                         & \multicolumn{2}{c}{$7.0 \times 10^{-3}$}   &  $0.88$  \\
    $X_\text{Ne}$\tablefootmark{c}                        & \multicolumn{2}{c}{$1.3 \times 10^{-3}$}   &  $1.0$   \\
    $X_\text{Mg}$\tablefootmark{b}                        & \multicolumn{2}{c}{$7.0 \times 10^{-4}$}   &  $1.0$   \\
    $X_\text{Al}$\tablefootmark{b}                        & \multicolumn{2}{c}{$7.0 \times 10^{-5}$}   &  $1.31$  \\
    $X_\text{Si}$\tablefootmark{b}                        & \multicolumn{2}{c}{$5.5 \times 10^{-4}$}   &  $0.75$  \\
    $X_\text{P}$\tablefootmark{b}                         & \multicolumn{2}{c}{$6.4 \times 10^{-6}$}   &  $1.0$   \\
    $X_\text{S}$\tablefootmark{b}                         & \multicolumn{2}{c}{$5.0 \times 10^{-4}$}   &  $1.0$   \\
    $X_\text{Cl}$\tablefootmark{c}                        & \multicolumn{2}{c}{$8.2 \times 10^{-6}$}   &  $1.0$   \\
    $X_\text{Ar}$\tablefootmark{c}                        & \multicolumn{2}{c}{$7.3 \times 10^{-5}$}   &  $1.0$   \\
    $X_\text{K}$\tablefootmark{c}                         & \multicolumn{2}{c}{$3.1 \times 10^{-6}$}   &  $1.0$   \\
    $X_\text{Ca}$\tablefootmark{c}                        & \multicolumn{2}{c}{$6.1 \times 10^{-5}$}   &  $1.0$   \\
    $X_\text{Fe}$\tablefootmark{b,d}                      & \multicolumn{2}{c}{$1.4 \times 10^{-3}$}   &  $1.0$   \\
            
  \hline
  \end{tabular}
  \tablefoot{
        \tablefoottext{a}{Ratio of mass fractions relative to their solar value from \citet{Asplund+2009}.}
        \tablefoottext{b}{Abundance taken from \citet{GimenezGarcia+2016}}
        \tablefoottext{c}{Solar abundance assumed, taken from \citet{Asplund+2009}}
        \tablefoottext{d}{Fe also includes  the iron group elements Sc, Ti, V, Cr, Mn, Co, and Ni.
                          See \citet{GKH2002} for relative abundances.}
  }  
\end{table}    

        Starting from the previously described non-HD model, the first hydrodynamically consistent model
        was calculated (first HD model). For this first model, no X-rays were included, so that we simulated the situation for
        an unperturbed B-star wind with stellar parameters similar to the donor of Vela X-1. This model is 
        not only helpful for comparisons, it is also essential for establishing the (electron) temperature stratification
        $T_\text{e}(r)$ for our follow-up models, since we left $T_\text{e}(r)$ unchanged when we included X-rays. Fixing
        $T_\text{e}(r)$ is of course a simplification since the influence of the X-rays, on the ionization structure, for example, might change
        the temperature stratification. Nonetheless, this also avoids overestimating the changes in $T_\text{e}(r)$ since our X-ray inclusion 
        is already rather approximate and not limited to a tiny NS source that would have a much smaller effect than our 1D treatment. 
        Even the
        first non-X-ray model yields a relatively low terminal wind velocity of $\varv_\infty = 532\,$km\,s$^{-1}$, thereby theoretically backing
        the empirically derived value of $\varv_\infty \approx 700\,$km\,s$^{-1}$ determined by \citet{GimenezGarcia+2016}. This
        is also in line with the results from \citet{Krticka+2012}, who obtained $\varv_\infty = 750\,$km\,s$^{-1}$ in their model without X-rays.
        These values are notably lower than what was inferred from IUE observations by \citet[$\sim$$1700\,$km/s]{Dupree+1980} and 
        \citet[$\sim$$1100\,$km/s]{Prinja+1990}, but as demonstrated in \citet{GimenezGarcia+2016}, models with high velocities would significantly 
        overestimate the observed UV line profiles. Furthermore, especially the population of \ion{C}{iv} is significantly larger than 
        for an unperturbed B star, resulting in a much stronger UV profile, as we discuss in Sect.\,\ref{sec:uvfeatures}.
        
        Based on the first HD model, two more HD models were calculated, adopting the previously discussed X-ray parameters corresponding
        to $L_\text{X} \approx 2.2 \cdot 10^{33}\,$erg/s and $L_\text{X} \approx 6.7 \cdot 10^{36}\,$erg/s, respectively. 
        We refer to the two cases as ``moderate'' and ``strong'' illumination test cases in the following. For the moderate case, we used the same X-ray
        onset radius as \citet{GimenezGarcia+2016}, namely $R_{0,1} \equiv d_0 = 1.2\,R_\ast$, while in the strong case, we used the X-rays from the moderate case
        plus an additional second component with an onset radius $R_{0,2} = d_\textsc{ns} = 1.8\,R_\ast$. To avoid multiple indices, we
        use the terms $d_0$ and $d_\textsc{ns}$ in all figures where these onset radii are outlined.
        
        Since \citet{GimenezGarcia+2016} used X-ray parameters that correspond to neither of these two cases, we also calculated a model with their
        X-ray parameters. However, these results were almost identical to our results when we used the $L_\text{X}$ that is motivated by eclipse measurements. 
        Therefore, we refrain from discussing this additional X-ray HD model here more explicitly. Nevertheless, it is worth mentioning
        that the X-ray implementation in PoWR (see Sect.\,\ref{sec:powrxray}) does not necessarily lead to the same value of $L_\text{X}$
        when the underlying atomic data and the clumping stratification are changed. While \citet{GimenezGarcia+2016} have 
        $L_\text{X} \approx 5 \cdot 10^{33}\,$erg/s, the HD model with the same $R_{0}$, $T_\text{X}$ , and $X_\text{fill}$ has $L_\text{X} \approx 8 \cdot 10^{34}\,$erg/s, meaning that the
        resulting X-ray luminosity is more than an order of magnitude higher. This is mainly an indirect effect of the slight 
        change in the clumping stratification, which leads to a different onset. In our models, clumping starts earlier, in particular at a mean density
        that is about two orders of magnitude higher than the density in \citet{GimenezGarcia+2016}. Since a significant part of the emerging X-ray flux in 
        our model stems from a wavelength regime where the wind is optically thin for X-rays, the higher clumping factor in the deeper layers becomes important and would have to
        be compensated for by a smaller X-ray filling factor if the same $L_\text{X}$ were to be kept. We instead decided to keep the 
        $X_\text{fill}$ from \citet{GimenezGarcia+2016} for the moderate case and assumed $X_\text{fill} = 2$ for the additional component
        in the strong case, while adjusting the X-ray temperatures $T_{X}$ to obtain the required values of $L_\text{X}$.

 \subsection{X-ray sensitive UV features}
   \label{sec:uvfeatures}

        Our first HD model without any X-rays yields $\varv_\infty \approx 532\,$km\,s$^{-1}$ together with $\log \dot{M} = -6.19$.
        As we show below, our two X-ray test cases affect the obtained terminal velocity, but leave the mass-loss rate almost unaffected.
        However, in all cases, the X-rays do significantly affect the ionization stratification in the wind and
        thus can have an imprint on the spectral lines that originate in the wind, most notably in the UV. 
        
\begin{figure}[th]
  \resizebox{\hsize}{!}{\includegraphics{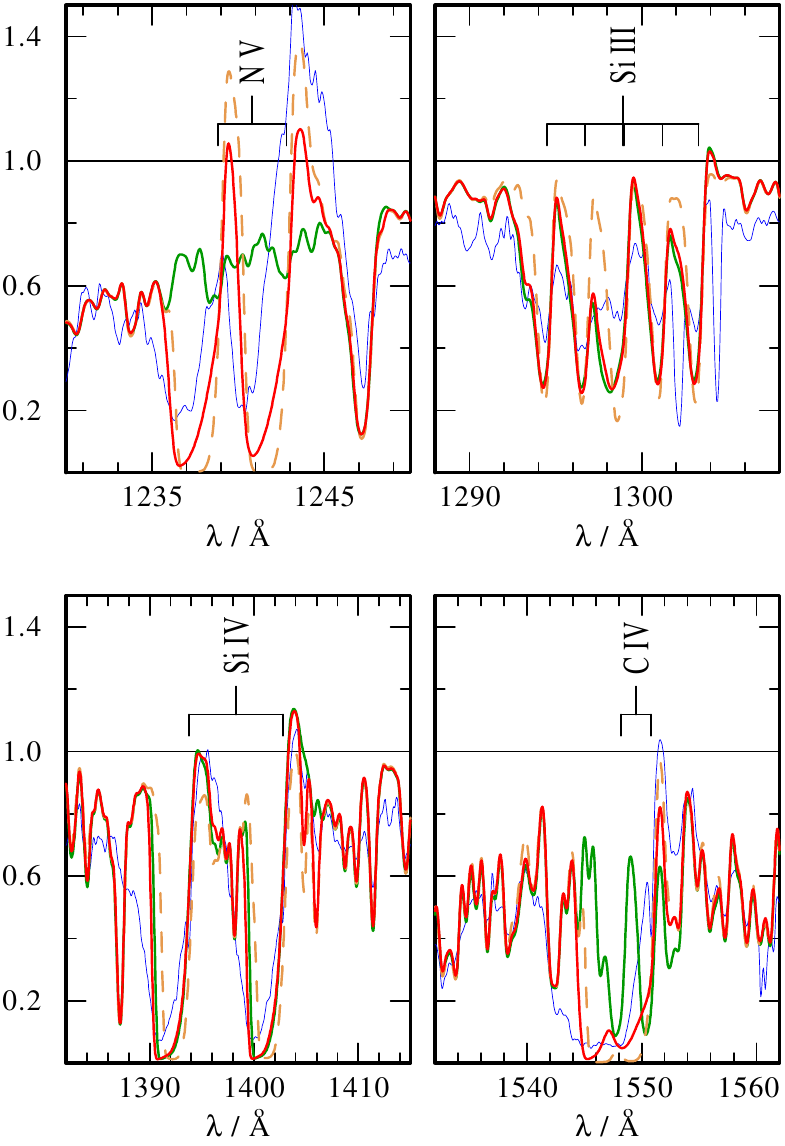}}
  \caption{Effect of including X-rays in the atmosphere calculations: The averaged observed normalized spectrum for
           selected UV lines (blue thin solid line) is compared to HD-PoWR models without (green solid line) as
                                         well as with moderate (red solid line) and strong X-ray illumination (brown dashed curve).}
  \label{fig:uvlines}
\end{figure}

\begin{figure}[th]
  \resizebox{\hsize}{!}{\includegraphics{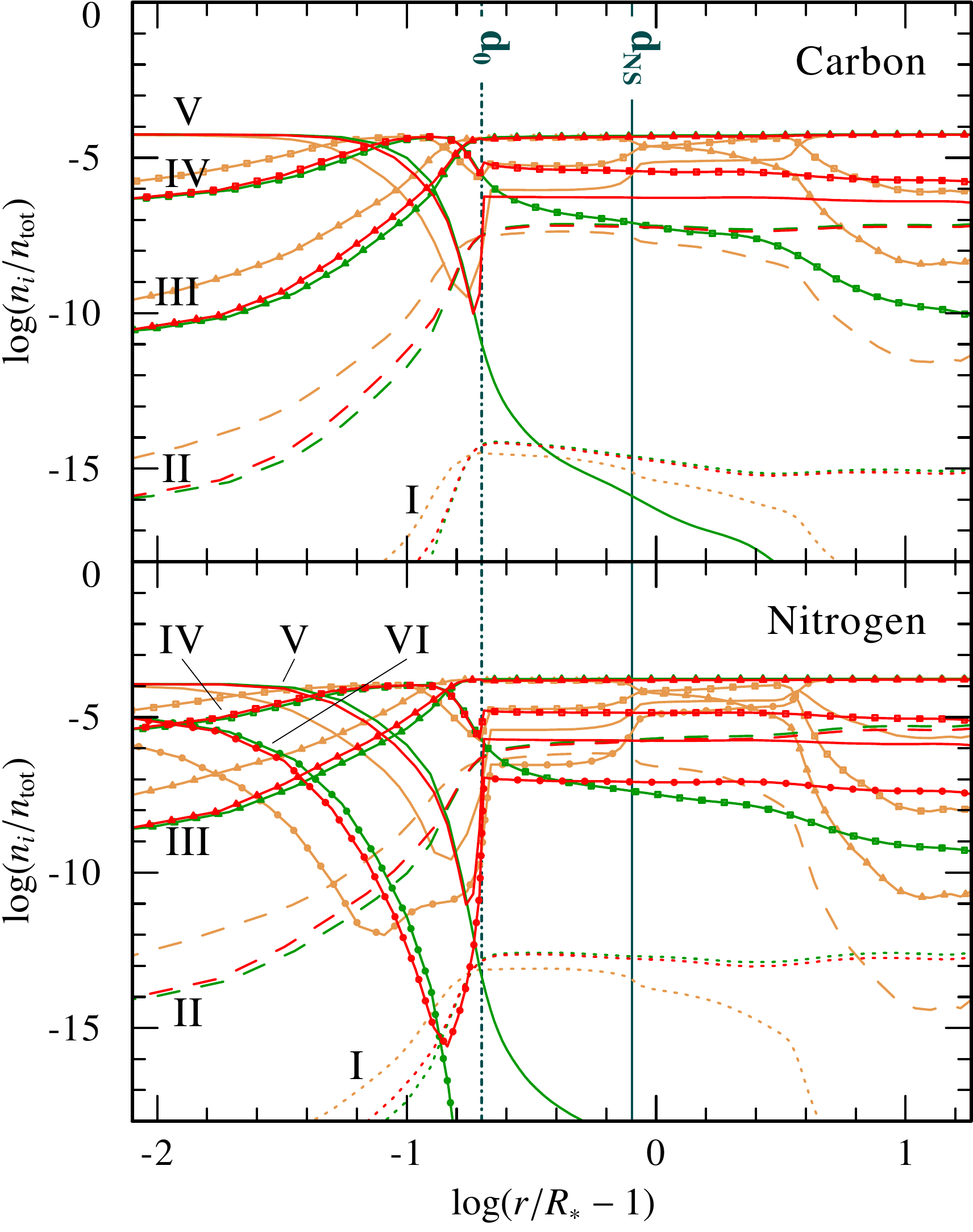}}
  \caption{Relative population numbers for the ground-state ion levels of carbon (upper panel) and nitrogen (lower panel)
           in the HD models with moderate $L_\text{X}$ (red curve), strong $L_\text{X}$ (brown dashed curve), and without X-rays (green curve).}
  \label{fig:popnums}
\end{figure}
        
        This is demonstrated in Fig.\,\ref{fig:uvlines} for a selection of UV lines for four different ions, where we compare the averaged
        observed IUE spectrum with our different HD consistent models. The observational data we used are identical to those used 
        by \citet{GimenezGarcia+2016}, where a more detailed description of the considered observations can be found.
        While \ion{Si}{iii} and \ion{Si}{iv}
        are essentially unaffected in the case of moderate X-ray illumination, the absorption trough of the \ion{C}{iv} doublet is significantly widened
        by an enhanced population of \ion{C}{iv} in the wind. For \ion{N}{v,} the effect is even more spectacular: Without X-rays,
				this ion is essentially absent in the outer wind, but the inclusion of X-rays offers an ionization source
        that then substantially populates \ion{N}{v}, even when using just the $L_\text{X}$ from the moderate illumination model.
        The idea that X-rays are responsible for this effect goes back to \citet{Cassinelli+1979} after the discovery of \ion{C}{iv}
        and \ion{N}{v} in the \textit{Copernicus} and \textit{Skylab} spectra of stars \citep{Snow+1976,Parsons+1979}, whose winds were 
        too cool for these ionization stages in radiative equilibrium. Since \citet{Cassinelli+1979} assumed a coronal X-ray
        origin, the proper accounting for X-rays that affect the ionization balance in the stellar wind was only performed in later studies \citep[e.g.,][]{Macfarlane+1993,Macfarlane+1994}
        despite the early discussions of donor wind changes caused by X-rays in the context of X-ray binaries \citep[e.g.,][]{BS1973,MH1975}.
        
        The changes in the ionization stages for our Vela X-1 donor wind model are visualized in Fig.\,\ref{fig:popnums}, where we compare the relative ground-state 
        populations for carbon and nitrogen.The leading ionization stage, that is the stage that most of the ions of an element populate, is also more or 
        less unaffected when considering only the $L_\text{X}$ from the moderate illumination model. The reason is that the population numbers in the leading stage,
        that is, \ion{C}{iii} and \ion{N}{iii} here, were several orders of magnitude larger than the those of \ion{C}{iv} or \ion{N}{v}
        in the model without X-rays, and the amount of X-rays in the moderate illumination case is just not high enough to significantly deplete the
        lower stages. The change in the ionization balance sets in immediately where the X-rays are inserted in the moderate illumination
        case, namely at $d_0$. 
        
        The whole ionization balance changes significantly for the model in which the strong $L_\text{X}$ is applied. Now the lower ionization
        stages become depleted outward of $d_\textsc{ns}$, where the additional X-ray component is taken into account. The depletion even 
        increases farther out at about $r \ga 2 d_\textsc{ns}$, where \ion{C}{v} and \ion{N}{vi} become the dominant ionization 
        stage. However, the X-ray flux is still not high enough to populate even higher stages, such as \ion{C}{vi} or \ion{N}{vii}. We 
        account for these high ionization stages in our calculations (cf. Table\,\ref{tab:datom}), but their population is so small that
        they are far below the scale in Fig.\,\ref{fig:popnums}.
        
        Since the strong illumination case consists of two X-ray components, setting in at $d_0$ and $d_\text{ns}$, we see corresponding
        changes in the ionization trend at these radii. The third change in the trend at $r \ga 2 d_\textsc{ns}$ does not reflect a
        change in the X-ray treatment, but instead indicates the region where the wind becomes transparent to X-rays. As the X-rays
        increasingly deplete the leading opacity sources such as \ion{He}{ii} or \ion{O}{iii} in outward direction, they essentially 
        remove the material that would be able to absorb X-rays. When these lower stages are fully depleted, the wind becomes transparent at X-ray 
        wavelengths and the leading ionization stages change significantly, for example, to \ion{He}{iii} or \ion{O}{vii}.
        
        The strong X-ray illumination also leaves quite a notable imprint on the UV lines in Fig.\,\ref{fig:uvlines}. The emission parts of
        \ion{C}{iv} and \ion{N}{v} lines become stronger and the blue edges of all P\,Cygni lines shrink, hinting at a lower wind
        velocity, which is indeed the case, as we show below when we discuss the stratification. Since the different amounts of X-ray
        illumination are a rough approximation of what we see from a system like Vela X-1 in different orbital phases, our resulting
        UV line profiles essentially mimic the so-called ``Hatchett-McCray effect'', that is, an orbital modulation of the UV lines 
        that is due to the
        change in position of the NS and its zone with higher ionization that is due to the X-rays. This was first discussed by 
        \citet{Hatchett+1977} and has later indeed been confirmed for Vela X-1 by comparing UV spectra from different orbital phases \citep[e.g.,][]{Kaper+1993,vanLoon+2001}.
        The change in the UV lines of our models also agrees qualitatively with the modeling results from \citet{vanLoon+2001}, who used
        a radiative transfer code based on the so-called ``Sobolev with exact integration'' (SEI) method \citep{Lamers+1987}. 
        In this work we do not aim to reproduce the precise shapes of the UV line profiles. For this task, optically thick wind clumping 
        in the radiative transfer calculations would also need to be considered, which is often discussed under the keywords
        ``macroclumping'' \citep[e.g.,][]{Oskinova+2007,Surlan+2013} or ``porosity'' \citep[e.g.,][]{Owocki2008,Sundqvist+2014} in physical and velocity space,
        from which we refrain here as it is beyond the scope of the present work.

 \subsection{Mass-loss rates}
   \label{sec:mdot}
        
    A compilation of the results from the hydrodynamically-consistent wind models for the donor star of Vela X-1 is presented in Table\,\ref{tab:modelparam}, 
	  where we list all three models and compare them to the empirical model with prescribed $\varv(r)$ by \citet{GimenezGarcia+2016}. The mass-loss rates of 
		all three HD models differ by less than $0.1\,$dex, while the terminal wind velocities vary as a result of the different X-ray illumination. While we 
		discuss the stratification details in the following section, we can conclude at this point that the X-rays do not strongly affect the layers of the critical point and below. 
		The radius of the critical point $r_\text{c}$ remains about the
    same for the moderate X-ray illumination case and moves only slightly inward for the strong illumination case, 
		resulting in a tiny increase in the mass-loss rate. The latter is exactly in line with the findings of \citet{MV1982} for
    HMXBs where the authors used a simpler Sobolev-based model. It also agrees with studies that have been performed for intrinsic X-rays in single-star winds
    \citep[e.g.,][]{KK2009}.
        
    Our value of the mass-loss rate in the non-X-ray model is $6.5 \cdot 10^{-7}\,\mathrm{M}_\odot/\mathrm{yr}$. This
    is slightly more than a factor of two lower than found by \citet{Krticka+2012}, namely $1.5 \cdot 10^{-6}\,\mathrm{M}_\odot/\mathrm{yr}$.
    However, their modeling assumed  a smooth wind, while we used a depth-dependent microclumping approach with $D_\infty = 11$.
    While a simple comparison by multiplying our $\dot{M}$ with $\sqrt{D_\infty} \approx 3.3$ yields $2.1 \cdot 10^{-6}\,\mathrm{M}_\odot/\mathrm{yr}$
    would provide a good agreement, this diagnostic is only valid for wind lines that scale with $\rho^2$. Moreover,
    \citet{Muijres+2011} discovered that for a microclumping approach, the mass-loss rate should even increase compared to the
    unclumped situation, although the recent CMF-based calculations by \citet{Petrov+2016} might question this. 
    The reasons that our mass-loss rate is lower than  that of \citet{Krticka+2012} likely lies in
    the different stellar parameters considered in the calculation. Even though both models use more or less the same $\log g$,
    the luminosity resulting from the parameters in \citet{Krticka+2012} is about $40\%$ higher, which likely propagates into
    the higher mass-loss rate.
        
    When we compare our result of $\log \dot{M} = -6.19$ to the predictions from \citet{Vink+2000,Vink+2001} assuming 
    solar metallicity, their prediction of $-5.06$ is more than an order of magnitude higher. For the stellar parameters from
    \citet{Krticka+2012}, the prediction would be $-5.61$, which is also higher than obtained by \citet{Krticka+2012}, but only by about
    $0.2\,$dex. The main reason for the larger discrepancy with the predictions from \citet{Vink+2000,Vink+2001} is
    our lower temperature of $T_{2/3} = 23.5\,$kK. This is located between the two bistability jumps identified in \citet{Vink+1999},
    while the $27\,$kK assumed by \citet{Krticka+2012} is already above the higher bistability jump implemented in their
    formula. The steep increase in mass-loss rates when transitioning to cooler temperatures in the calculations from
    \citet{Vink+1999} is not seen in our model. However, the observed bistability according
    to \citet{Lamers+1995} should occur around $21\,$kK, and the recent calculations from \citet{Petrov+2016} also found it
    in a similar range, which would place even our value of $T_{2/3}$ still on the hotter side. Unfortunately, \citet{Petrov+2016}
    does not offer models with $20\,\mathrm{M}_\odot$ around our $T_{2/3}$, which would have allowed for a direct comparison of $\dot{M}$.
    In any case, the donor star of an HMXB is a particular situation and although most abundances in our model are assumed to be solar, 
    there is a significant nitrogen enrichment and carbon 
    depletion as derived by \citet{GimenezGarcia+2016}, which we take into account, so the comparison with models based on scaled 
    main-sequence compositions is certainly limited. Nevertheless, our models provides a first interesting test case for the bistability
    jump regime, which we discuss in the context of the driving ions in Sect.\,\ref{sec:driving} in more detail.

 \subsection{Stratification}
   \label{sec:strat}
        
\begin{figure}[t]
  \resizebox{\hsize}{!}{\includegraphics[angle=270]{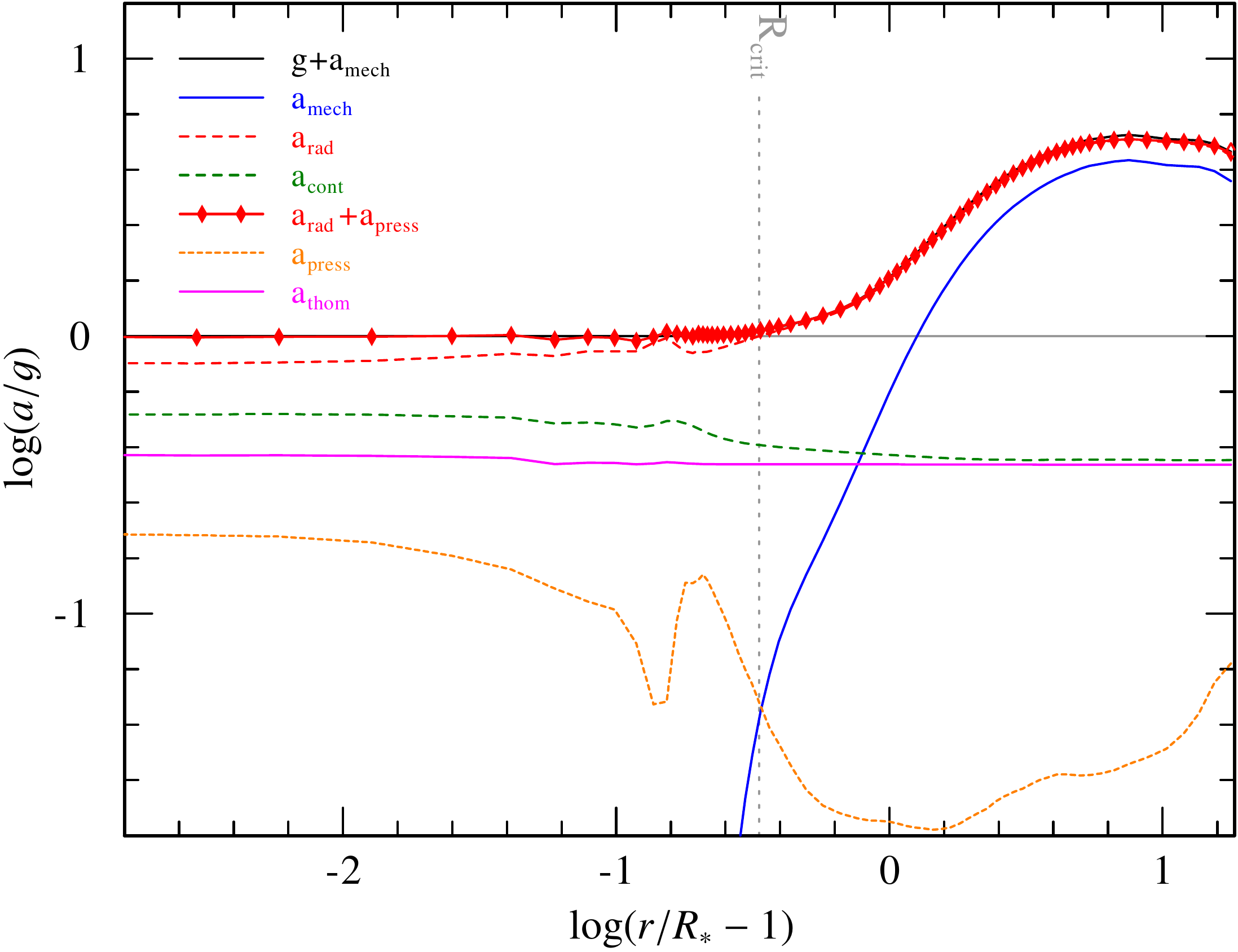}}
  \caption{Detailed acceleration stratification for the HD consistent 
           model using the moderate X-ray flux. The wind acceleration (thick red diamond line) is 
         compared to the repulsive sum of inertia and gravitational acceleration $g(r)$~(black
           line). The input parameters of the model are given in Table\,\ref{tab:modelinput}, while the resulting
         quantities are listed in Table\,\ref{tab:modelparam}. In order to properly handle the
         various scales, this is a double-logarithmic plot with all acceleration terms normalized to $g(r)$.}
  \label{fig:acc-eclipse}
\end{figure}

    In an HD PoWR model, the outward and inward forces balance each other throughout
    the whole atmosphere, thus providing a self-consistent stratification. A visual check for the success of the solution method 
    can be made by plotting the different accelerations, namely the total radiative acceleration $a_\text{rad}(r)$, the 
    acceleration from gas pressure as a result of temperature and turbulence $a_\text{press}(r)$, the gravitational
    acceleration $g(r)$, and the inertia $a_\text{mech}(r) = \varv(r) \frac{\mathrm{d}\varv}{\mathrm{d}r}$. This is shown in
    Fig.\,\ref{fig:acc-eclipse} for the model accounting for moderate X-ray illumination. The sum of $a_\text{rad}$
    and $a_\text{press}$ matches the sum of $g$ and $a_\text{mech}$ throughout the atmosphere, and thus our model is indeed
    hydrodynamically consistent.

Inspection of the contributions from the different accelerations in Fig.\,\ref{fig:acc-eclipse} reveals that the
general picture is very similar to what we obtained for the significantly hotter O supergiant in \citet{Sander+2017}. 
In the wind, only the line acceleration and Thomson scattering are important, while in
the inner subsonic regime, the gas pressure and the contributions from the continuum opacities from bound-free transitions
have also to be considered. However, the curve shapes differ in detail, and the 
increase of $a_\text{rad}$ beyond the critical point is significantly shallower than in the case of the O supergiant \citep[cf.~Fig.\,6 in][]{Sander+2017}. This is due to 
the different ions that contribute in the temperature regime of an early-B supergiant with $T_\ast = 25.5\,$kK compared to the $42\,$kK of the
O star discussed in \citet{Sander+2017}. The detailed elemental contributions to the driving are discussed in Sect.\,\ref{sec:driving}.

\begin{figure}[th]
  \resizebox{\hsize}{!}{\includegraphics{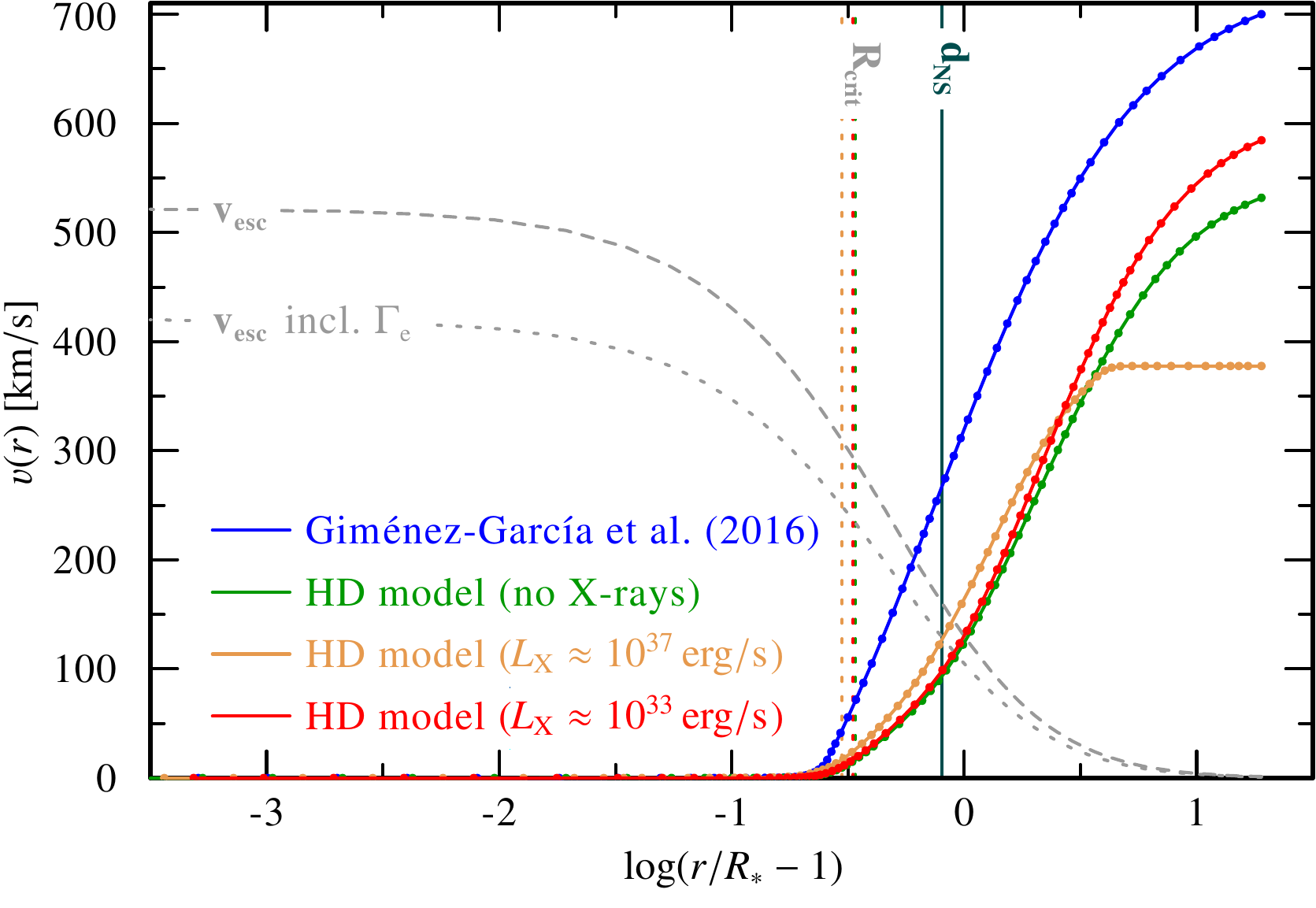}}
  \caption{Velocity field for Vela X-1 from \citet{GimenezGarcia+2016} compared to 
           the fields from the HD consistent models presented in this work using different
           X-ray fields. The orange line marks the location of the NS at
         a radius of $d_\textsc{ns} = 1.8\,R_\ast$ from the center of the donor, while the
         colored dashed lines denote the location of the critical point in the corresponding
         HD models. The grey dashed lines denote the escape speed $\varv_\text{esc}(r)$
                                 without (large dashes) and with (small dashes) accounting for the reduction due 
                                 to $\Gamma_\text{e}$.}
  \label{fig:velocmp}
\end{figure}

\begin{figure}[th]
  \resizebox{\hsize}{!}{\includegraphics[angle=270]{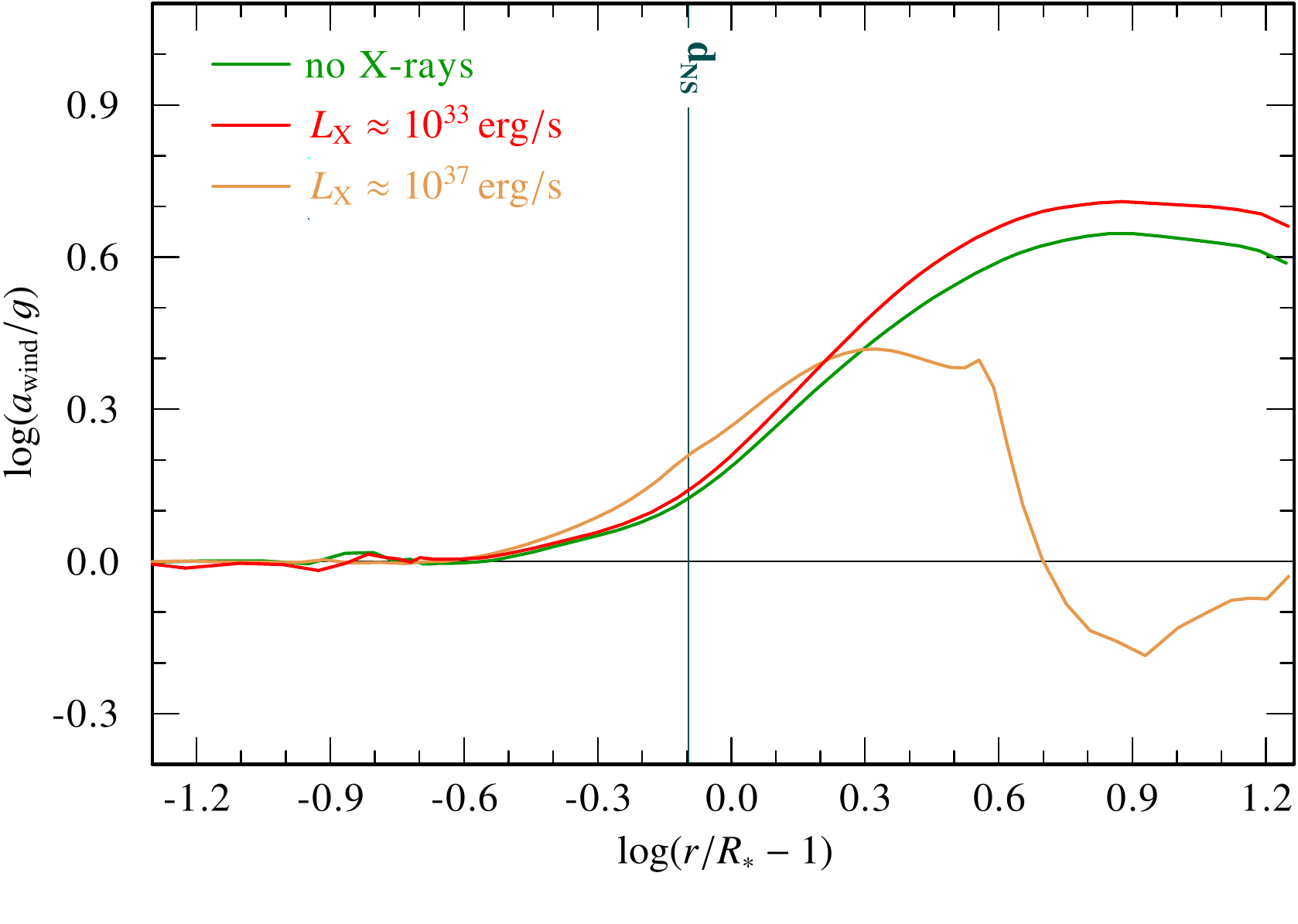}}
  \caption{Comparison of the total wind acceleration -- normalized to $g(r)$ -- for the three HD consistent 
           models using different amounts of X-ray illumination.}
  \label{fig:acc-cmp}
\end{figure}

A comparison of the velocity field $\varv(r)$ from \citet{GimenezGarcia+2016} using a prescribed $\beta$-law connected to a 
consistent hydrostatic solution \citep[see][for technical details]{Sander+2015} and from our HD consistent model is shown in Fig.\,\ref{fig:velocmp}. 
While we see a sharp increase for $\varv(r)$ in the model from \citet{GimenezGarcia+2016} around and outward of the sonic point, which marks our 
critical point in the HD case, the increase is more moderate for the self-consistent HD models.
A similar situation occurred for the O supergiant model in \citet{Sander+2017} and is likely related to two things: first,
the $\beta$-value used for the prescribed law is -- if not simply adopted but motivated by observations -- typically inferred from H$\alpha$.
Second, $\varv(r)$ in the region around the connection point between the quasi-hydrostatic layers and the $\beta$-law regime can
significantly violate the acceleration balance we aim at with our HD models. When these models are used to obtain empirical stellar and wind parameters, 
this is usually not a problem, but as soon as conclusions are to be drawn from the detailed stratification, this can lead to errors, most 
notably in this transition regime. 

A closer look reveals that all HD solutions behave rather similar in the inner wind near the critical point, but in the outer layers, 
the effects of the different amount of X-rays become clearly noticeable. In the model with moderate X-ray illumination, the amount of X-rays is just enough
to ionize the wind such that the population of some of the driving ions such as \ion{N}{v} increases and additional driving is provided. However,
when the amount of X-rays becomes too high, as we see in the strongly X-ray illuminated model, there is so much ionization that important driving ions are depopulated,
causing a sharp decrease in the line acceleration in the outer wind. This effect is also illustrated in Fig.\,\ref{fig:acc-cmp}, where we compare the wind accelerations
for all three HD solutions. Here, we further note that the flattening of the velocity field for the strong illumination case seen in Fig.\,\ref{fig:velocmp} is an artifact that is due to our technical limitations to monotonic velocity fields in the CMF radiative transfer rooted in the structure of the frequency
boundary together with the applied elimination scheme \citep[for more details, see][especially their Appendix A]{Mihalas+1975}. Since the normalized wind acceleration drops below unity here, we would 
have a deceleration in reality and thus an even lower terminal velocity than obtained in this work, which we cannot model because
of the limitation to monotonic velocity fields.
Interestingly, \citet{Kaper+1993}  suggested a have-monotonic velocity field for the donor wind of Vela X-1 because they had difficulties to model the changes of the UV lines
when assuming a monotonic $\varv(r)$, but they discussed this with regard to wind-intrinsic instabilities and not wind deceleration that is due to X-ray ionization. 
In the work of \citet{Krticka+2012}, the velocity in the direction toward the NS stops to increase much farther inward than in our model. The reasons
are hard to determine, since not only their X-ray treatment,  but also their radiative transfer treatment are entirely different. In their X-ray models, they use the
Sobolev line acceleration corrected with factors obtained from their CMF-based calculation in the non-X-ray case. They furthermore do not seem to consider the acceleration
due to Thomson scattering, which -- unlike the line acceleration -- does not break down in our models.
In any case, Fig.\,\ref{fig:velocmp} demonstrates that in all models the wind is accelerated to velocities higher than the local escape speed,
allowing the material to leave the star even if no further acceleration were to occur. Interestingly, for Vela X-1, this is coincidentally just around or slightly beyond
the radius where the NS is located. A rather slow wind at this distance fosters the accretion of wind material by the compact object, as we
discuss in more detail in Sect.\,\ref{sec:accretion}.
Furthermore, in the strong X-ray illumination model, the wind speed does not surpass the escape speed at $R_\ast$ of $\varv_\text{esc} = 521\,$km/s, even when we consider the reduction due to $\Gamma_\text{e}$, which would decrease the escape speed to $420\,$km/s. In the other two cases, we obtain a ratio $\varv_\infty/\varv_\text{esc} \approx 1.3$ when we account for $\Gamma_\text{e}$ in $\varv_\text{esc}$. This ratio is typical for the regime between the two bistability jumps \citep[e.g.][]{Lamers+1995,Vink+1999}, which is especially interesting as our mass-loss rate would be associated with the regime above the jumps.  

Although the radiative acceleration drops for $r \ga 2 d_\textsc{ns}$, Fig.\,\ref{fig:acc-cmp} also illustrates that it does not vanish completely, and the wind might therefore not be shut off completely, even in the strongly ionized region. Our quite approximate X-ray 
treatment  and the fixed temperature structure might be a caveat here, but it is noteworthy that the large breakdown of the acceleration 
and thus the strongest effect of the X-rays does not occur at the distance of the NS, but instead much farther outside for $r \gtrsim 3\,R_\ast \approx 2 d_\textsc{ns}$ , where
the stratification also becomes optically thin for X-ray wavelengths and the ionization balance changes, as discussed in Sect.\,\ref{sec:uvfeatures}.
A more sophisticated treatment of the situation is needed to verify these results, but this might have interesting consequences for
the proper wind treatment in multidimensional time-dependent HD simulations of HMXBs. 

\begin{figure}[t]
  \resizebox{\hsize}{!}{\includegraphics[angle=270]{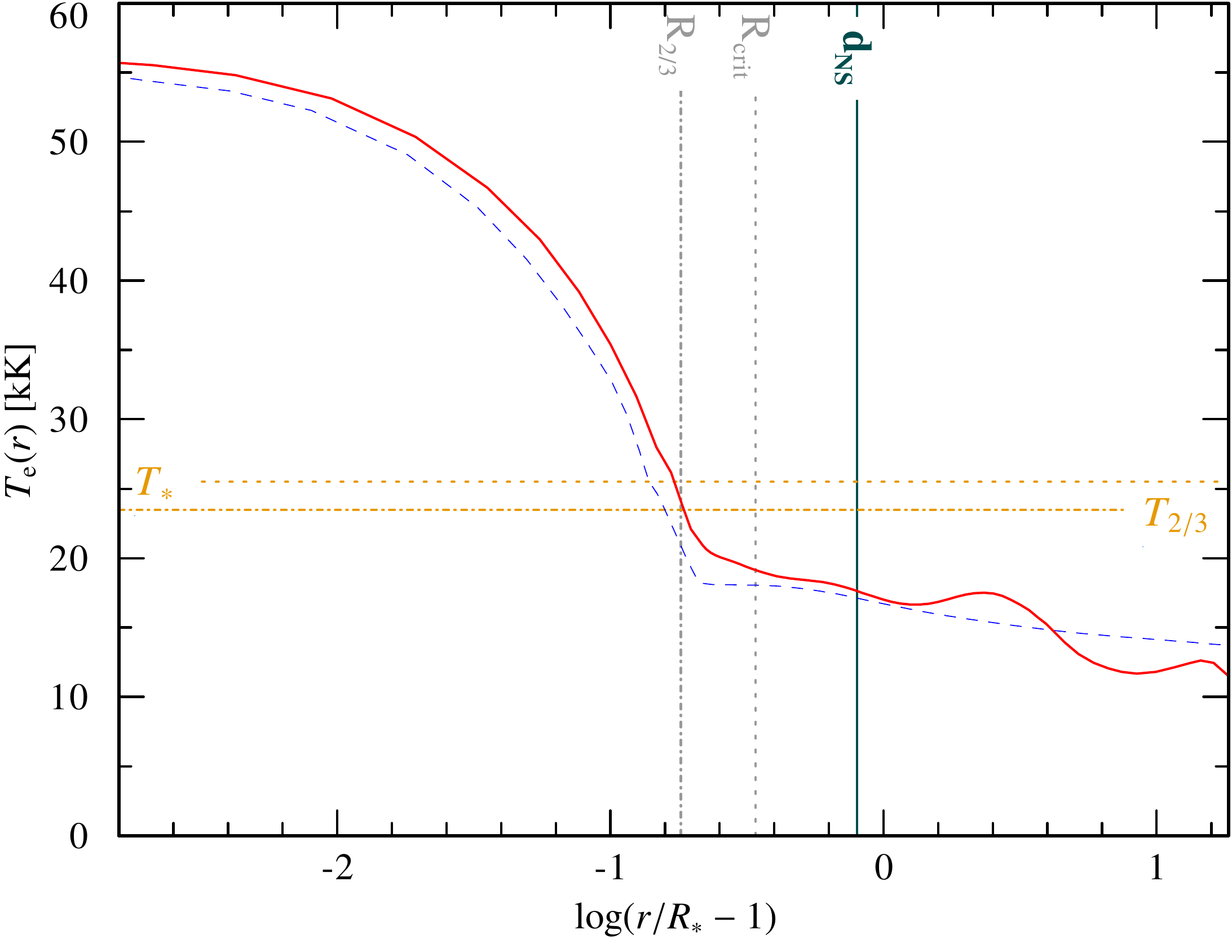}}
  \caption{Electron temperature stratification for the HD consistent models (red solid curve)
           compared to those of a model with a prescribed wind stratification (blue dashed curve)
         based on \citet{GimenezGarcia+2016}. Horizontal lines mark $T_\ast$ and $T_{2/3}$ for the
         HD-models. Vertical lines denote $R_{2/3}$ as well as the critical radius for the HD-model
         without X-rays and the distance of the NS.}
  \label{fig:temperature}
\end{figure}

The (electron) temperature stratification $T_\text{e}(r)$ for all HD models is displayed in Fig.\,\ref{fig:temperature}, where we also
show the stratification from \citet{GimenezGarcia+2016} for comparison. The results do not differ much, but the HD model is slightly hotter in the inner part and has minor non-monotonic parts in the outer wind. The shift in the
optically thick part likely arises from the lower microturbulent velocity assumed in the HD calculations, which we fixed in the main
iteration at the more ``canonical'' $10\,$km/s for B-supergiants \citep[e.g.,][]{LH2007} compared to $30\,$km/s in \citet{GimenezGarcia+2016}. 
Furthermore, the region between $R_\text{crit}$
and $R_{2/3}$ is smoother, which likely results from the fact that the HD-stratification avoids artifacts that can 
arise when the velocity fields of the subsonic and the wind regime
are connected. We note that both studies neglected the changes in
the electron temperature structure that are due to X-rays.

\begin{table}
  \caption{Results from the HD consistent model for the Vela X-1 donor star HD\,77581}
  \label{tab:modelparam}
  \centering
  \begin{tabular}{l c ccc}
  \hline\hline
     Quantity                 &  \! GG2016\tablefootmark{a}\!\! & \multicolumn{3}{c}{\hrulefill~HD models~\hrulefill} \\
                              &     {\small empirical}      &  {\small no $L_\text{X}$} &  \!\!\!{\small ``moderate''}\tablefootmark{b}\!\!\!   &  {\small ``strong''}\tablefootmark{c}       \\
        \hline
                $\log\,(T_\text{X}\,[\text{K}])$ \rule[0mm]{0mm}{3mm}           & $7$ &  --     &  $6.6$ & $7, 8.9$ \\
                $X_\text{fill}$                                       & 0.05   &  --  &   0.05  &  0.05, 2.0  \\
                $R_\text{0}$\,[$R_\ast$]                              & 1.2  &   --   &   1.2   &  $1.2, 1.8$ \\
                  
    \medskip
    $\log\,(L_\text{X}\,[\text{erg\,s}^{-1}])$\!\!\!\!       & 33.7 &   --   &   33.4  &   36.9  \\

    $R_\text{crit}$\,[$R_\ast$]                              &   --   &   1.34 &  1.33  &    1.30  \\
    \textbf{$\mathbf{\boldsymbol{\log}\,(\dot{M}}$\,[$\boldsymbol{M_\odot}\,\text{yr}^{-1}$]$)$}\!\!\!\!\!\!\!\!  &  -6.20  &  -6.19  &  -6.18  &  -6.07 \\
    \medskip
    \textbf{$\boldsymbol{\varv_\infty}$\,[km\,s$^{-1}$]}          &   700  &   532  &  584   &    378   \\                 

    $\log\,(R_\text{t}\,[\text{R}_\odot])$\tablefootmark{d}       &   2.20 &   2.12 &    2.14 &    1.94  \\
    $\log\,(Q_\text{ws}\,[\text{cgs}])$\tablefootmark{d}          & -12.12 & -11.94 &  -11.99 &  -11.60  \\
    $T_{2/3}$\,[kK]                                       &  24.4  &  23.5  &   23.5   &   23.8   \\
                $\log\,(g_{2/3}$\,[cm\,s$^{-2}$])                    &  2.79  &   2.69 &   2.70  &    2.72  \\
    $q_\text{ion}$\tablefootmark{e}                       &   0.77 &   0.75 &   0.75  &    0.77  \\
    \medskip
    $\Gamma_\text{e}$                                     &   0.34 &   0.35 &  0.35  &    0.36  \\

    $\eta = \dot{M} \varv_\infty c / L$                   &   0.07 &   0.06 &  0.06  &    0.05  \\                 
    $\log D_\text{mom}$\tablefootmark{f}
                                                          &  28.7  &  28.6  & 28.6   &   28.6   \\                 
  \hline
  \end{tabular}
  \tablefoot{
      \tablefoottext{a}{Empirical values derived by \citet{GimenezGarcia+2016}}
      \tablefoottext{b}{moderate X-ray illumination of the donor wind with $L_\text{X}$ corresponding to eclipse measurement}
      \tablefoottext{c}{strong X-ray illumination of the donor wind with $L_\text{X}$ corresponding to an unobscured NS situation}
            \tablefoottext{d}{\textit{\textup{Transformed radius}} $R_\text{t}$ and \textit{\textup{wind strength parameter}} $Q_\text{ws}$, see Sect.\,4.2 of \citet{Sander+2017} for definitions and model usage}
                        \tablefoottext{e}{Mean ionization parameter $q_\text{ion}$, as introduced in \citet{Sander+2015}}
      \tablefoottext{f}{Modified wind momentum in units of g\,cm\,s$^{-2}$, defined as $D_\text{mom} = \dot{M} \varv_\infty \sqrt{R_\ast/R_\odot}$ \citep[see, e.g.,][]{KP2000}}
  }  
\end{table} 
  
  The stratifications from all three HD consistent wind models are provided as tables in Appendix \ref{sec:apptables}.

  \subsection{Accretion estimation}
        \label{sec:accretion}

        In the so-called ``wind-fed HMXBs'', the wind of the donor star is accreted by the compact object, in our case, an NS star. 
        The empirical results from \citet{GimenezGarcia+2016} place Vela~X-1 in the so-called ``direct accretion regime'', using both the
        $\varv_\text{wind}$-$P_\text{spin}$ and the $\varv_\text{wind}$-$\dot{M}$ planes to visualize the scheme from \citet{Bozzo+2008}. 
        This scheme compares the relative positions of the NS magnetospheric radius with its accretion and corotation radii. The equations
        for transitions between different accretion regimes can be expressed with the help of the donor's wind parameters.
        Since our results for the mass-loss rate in general confirm the findings of \cite{GimenezGarcia+2016} and the wind velocity
        at the distance of the NS is even slightly lower than the one inferred from their prescribed law,
        the assumption that Vela~X-1 is set in the direct accretion regime is corroborated by our results. For this case, we can
        expect the X-ray luminosity $L_\text{X}$ to be roughly on the order of the accretion luminosity $L_\text{acc}$, which
        we estimate through Bondi-Hoyle-Lyttleton (BHL) accretion \citep{HL1939,BH1944,DO1973}. A more detailed discussion of BHL accretion 
        and the different accretion regimes can be found in the recent review by \citet{HMXBReview2017}.
        
  The Bondi-Hoyle radius or ``accretion radius''
  \begin{equation}
     \label{eq:racc}
     R_\text{acc} = \frac{2 G M_\textsc{ns}}{\varv_\text{rel}^2}
  \end{equation}
        describes the radius around the NS within which wind material can be captured and accreted. $M_\textsc{ns}$
        denotes the mass of the NS, and $G$ is the gravitational constant. The relevant velocity
        for $R_\text{acc}$ is the relative velocity $\varv_\text{rel}$ between the NS and the donor wind, resulting
        from the radial wind velocity $\varv(r=d_\textsc{ns})$ and the orbital velocity of the NS, $\varv_\text{orb}$. For a
        circular orbit, which is quite a good approximation for the case of Vela X-1 ($e \la 0.1$, see Table\,\ref{tab:sysparams}), the
        vector calculation simplifies to
        \begin{equation}
          \label{eq:vrel}
          \varv_\text{rel} = \sqrt{\varv_\text{orb}^2 + \varv^2(d_\textsc{ns})}\text{.}
        \end{equation}
        When estimating the relative velocity, a common assumption is to use $\varv_\text{rel} \approx \varv(d_\textsc{ns})$ assuming that the orbital velocity is
        much lower than the wind velocity ($\varv_\text{orb}^2 \ll \varv^2(d_\textsc{ns})$). 
        Since the shape of the velocity field is usually unknown, assuming $\varv(d_\textsc{ns}) \approx \varv_\infty$ is
        sometimes used as well, which would allow us to replace $\varv_\text{rel}$ with $\varv_\infty$ in Eq.\,(\ref{eq:racc}). However, both assumptions
        cannot be taken for granted for a particular system and thus can lead to significant errors in the accretion estimation if used inadvertently. Typical 
        orbital separations are $d_\textsc{ns} \approx 2\,R_\ast$ , at which it is by no means guaranteed that the wind has already reached its terminal velocity.
        
        In the case of Vela X-1, the NS seems to be located even slightly closer. Using the velocity field from our model with moderate X-ray illumination, we 
        predict a value of $\varv(d_\textsc{ns})$ that is almost an order of magnitude lower than $\varv_\infty$, meaning that this would be causing a huge error due the steep 
        velocity dependence, as we show in the further calculation. For other systems, this error might be smaller, but overestimating $\varv_\text{rel}$ by a 
        factor of $2$ would be rather typical when assuming $\varv(d_\textsc{ns}) \approx \varv_\infty$. Second, the low wind velocity at the distance of 
        the NS also means that
        the assumption $\varv_\text{orb}^2 \ll \varv^2(d_\textsc{ns})$ is not true.With the prescribed velocity field, \citet{GimenezGarcia+2016} have reported 
        $\varv_\text{orb} \approx \varv(d_\textsc{ns})$. The HD consistent solution now predicts that the wind velocity at the location of
        the NS is even lower than the orbital speed of the compact object. Thus Eq.\,(\ref{eq:vrel}) yields in our case $\varv_\text{rel} \approx 300\,$km\,s$^{-1}$.
        
        Assuming that the potential energy from the accreted matter is completely converted into X-rays, we obtain the accretion luminosity,
  \begin{equation}
    \label{eq:laccdef}
    L_\text{acc} = \frac{G M_\textsc{ns}\dot{M}_\text{acc}}{R_\textsc{ns}}
 ,\end{equation}
        with $R_\textsc{ns}$ denoting the radius of the NS and $\dot{M}_\text{acc}$ the mass accretion rate. Assuming direct wind accretion, the latter can be 
        expressed as
  \begin{equation}
    \label{eq:maccstart}
    \dot{M}_\text{acc} = \zeta \pi R_\text{acc}^2 \varv_\text{rel} \rho(d_\textsc{ns})
 ,\end{equation}
        where $\zeta$ is a numerical factor introduced to correct for radiation pressure and finite gas cooling. For moderately luminous X-ray sources, this is 
        commonly taken as $\zeta\equiv 1$. Approximating the density $\rho(d_\textsc{ns})$ with the density from a stationary spherically symmetric wind, we can write
  \begin{equation}
    \rho(d_\textsc{ns}) = \frac{\dot{M}_\text{donor}}{4 \pi d_\textsc{ns}^2 \varv(d_\textsc{ns})}\text{.}
  \end{equation}
        This allows us to rewrite Eq.\,(\ref{eq:maccstart}) such that
        we can express the accretion rate with the wind mass-loss rate of the donor star, 
  \begin{equation}
      \label{eq:maccfraction}
      \dot{M}_\text{acc} = \zeta \frac{R_\text{acc}^2}{4 d_\textsc{ns}^2} \frac{\varv_\text{rel}}{\varv(d_\textsc{ns})} \dot{M}_\text{donor}\text{.}
  \end{equation}
        Now plugging Eq.\,(\ref{eq:maccfraction}) into Eq.\,(\ref{eq:laccdef}) and replacing $R_\text{acc}$ with the help of Eq.\,(\ref{eq:racc}), we obtain
    \begin{align}
      L_\text{acc} &= \frac{G M_\textsc{ns}}{R_\textsc{ns}} \dot{M}_\text{acc} \\
         \label{eq:Lxacc}
                            &= \zeta \frac{G M_\textsc{ns}}{R_\textsc{ns}} \frac{R_\text{acc}^2}{4 d_\textsc{ns}^2} \frac{\varv_\text{rel}}{\varv(d_\textsc{ns})} \dot{M}_\text{donor} \\
                            &= \zeta \frac{\left(G M_\textsc{ns}\right)^3}{R_\textsc{ns}} \frac{\dot{M}_\text{donor}}{d_\textsc{ns}^2~\varv_\text{rel}^3~\varv(d_\textsc{ns})}\text{,}
    \end{align}
        which eventually allows us to estimate $L_\text{acc}$ using our results. Applying typical values for the 
        NS ($M_\textsc{ns} = 1.4\,M_\odot$, $R_\textsc{ns} = 12\,$km, $d_\textsc{ns} = 1.8\,R_\ast$, $\varv_\textsc{ns} = 281\,$km\,s$^{-1}$) 
        and inferring a value of $\varv(d_\textsc{ns}) \approx 100\,$km\,s$^{-1}$ from our HD model, which 
        we essentially find regardless of the particular X-ray illumination, we find
    \begin{equation}
      L_\text{acc} \approx \zeta \cdot 6.5 \dots 8.7 \cdot 10^{37}\,\mathrm{erg/s}
   ,\end{equation}
        with the small range spanned by the mass-loss rates derived from no (lower value) and strong (higher value) X-ray illumination.
        This is on the order of what we used as $L_\text{X}$ in our strong illumination test case. Owing to the lower value of $\varv(d_\textsc{ns})$ compared to \citet{GimenezGarcia+2016},
        our value of $L_\text{acc}$ is almost an order of magnitude higher, resulting from the fact that $L_\text{acc}$ roughly scales with the inverse of this quantity to the fourth power.
        Although the average X-ray luminosity $\langle L_\text{X} \rangle \simeq 4.5 \cdot 10^{36}\,\mathrm{erg/s}$ \citep{Sako+1999,Fuerst+2010} is 
        also an order of magnitude lower, our estimate is
        still remarkably consistent given that we assumed accretion to be so efficient that all energy is converted into X-ray luminosity. 
        Since the BHL estimate is likely an upper limit, as indicated by simulations performed for accretion in binaries \citep[e.g.,][]{Theuns+1996}, 
        it is common to introduce an accretion efficiency parameter $\eta^\text{acc}_\text{eff} = L_\text{X}/L_\text{acc}$ -- sometimes also termed 
	      transformation factor -- which is typically assumed to be around $\eta^\text{acc}_\text{eff} \approx 0.1\dots0.3$ \citep[e.g.,][]{Negueruela2010,Oskinova+2012}.
        In our case, a value of $0.1$ would lead to an excellent agreement between our estimate and the observed $\langle L_\text{X} \rangle$.

  \subsection{Wind driving and the bistability jump}
          \label{sec:driving}        
        
\begin{figure*}[p!]
  \includegraphics[angle=0,width=0.95\textwidth]{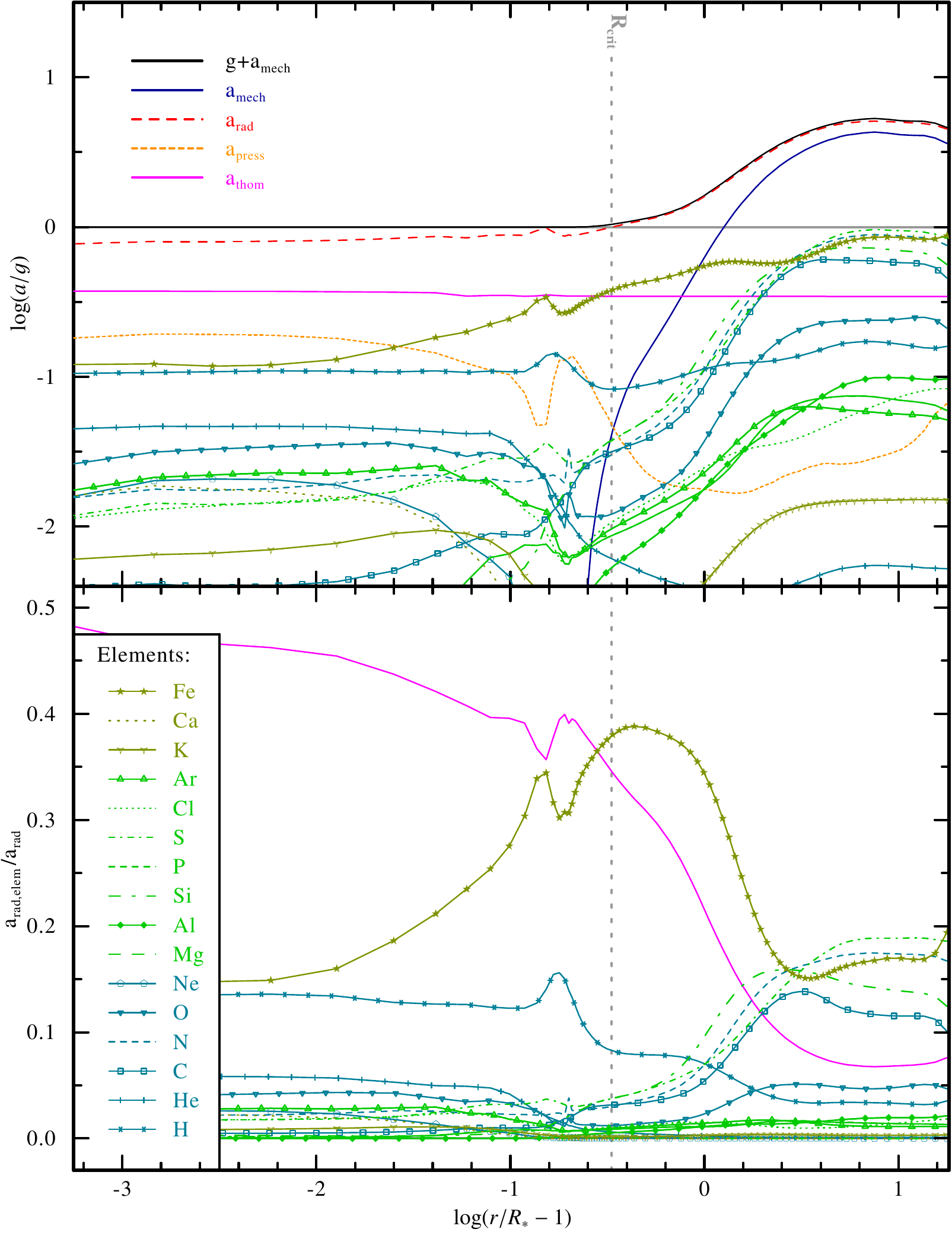}
  \caption{Absolute (upper panel) and relative (lower panel) contributions to the radiative
           acceleration from the different elements considered in the HD consistent
           atmosphere model incorporating the moderate X-ray flux. The total radiative acceleration and the acceleration 
           due to gas pressure are also shown in the upper panel for comparison. The lower panel shows the fraction
           that electron scattering (pink solid curve), and the various elements contribute to the 
           total radiative acceleration.}
  \label{fig:eclipseelem}
\end{figure*}

\begin{figure*}[ht]
  \resizebox{\hsize}{!}{\includegraphics{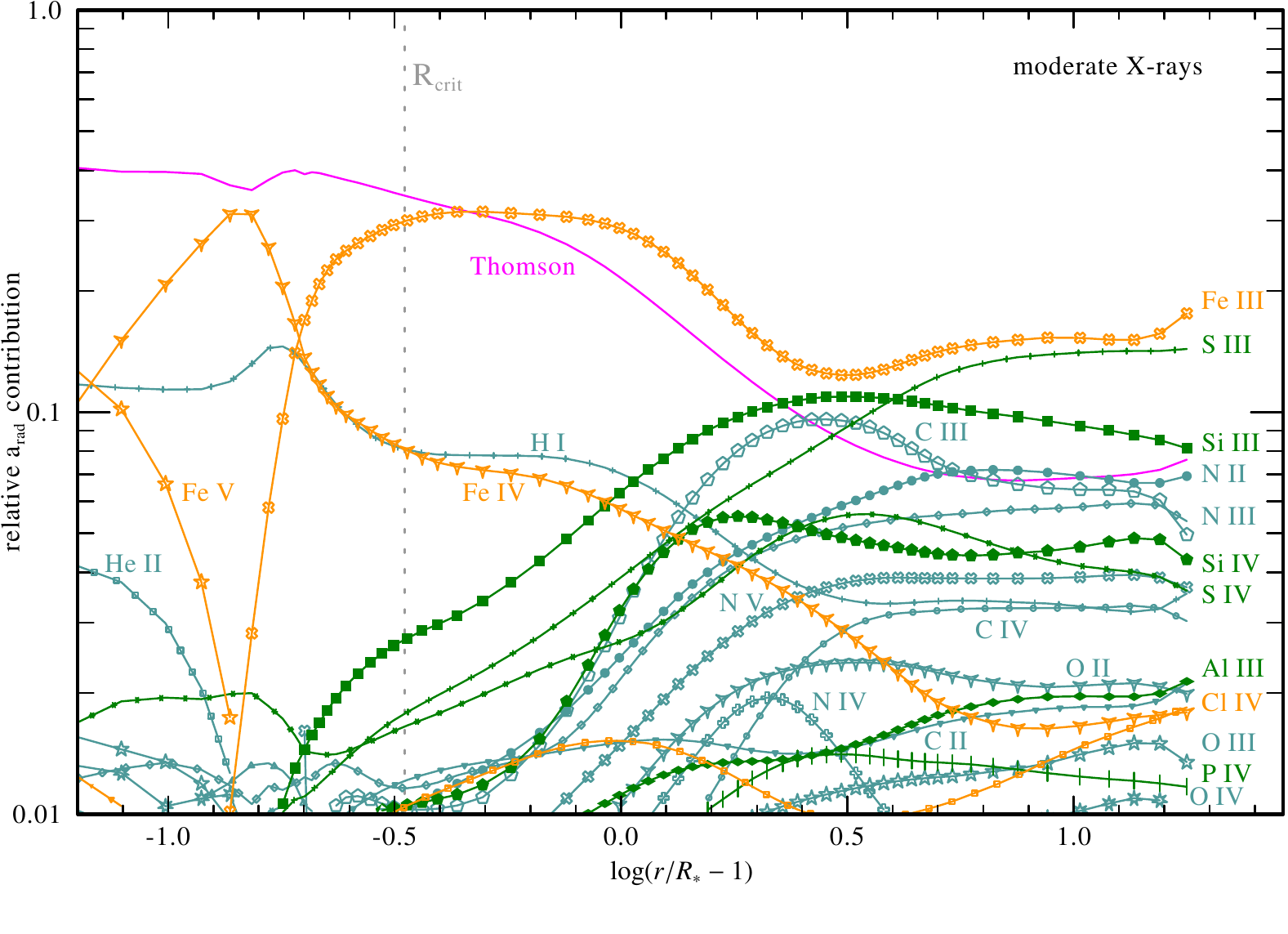}}
  \caption{Relative contributions to the radiative acceleration from the different ions (colored curves with various symbols) 
                 and due to scattering by free electrons (pink solid curve) for the HD consistent model with
                                         moderate X-ray illumination (cf.~Table\,\ref{tab:modelparam}). 
                                         Only ions that contribute at least 1\% to the radiative
                                         acceleration are shown. For better visibility of the various contributions on a percentage level, the y-axis is
                                         shown on a logarithmic scale.}
  \label{fig:leadions}
\end{figure*}

\begin{figure*}[htb]
  \resizebox{\hsize}{!}{\includegraphics{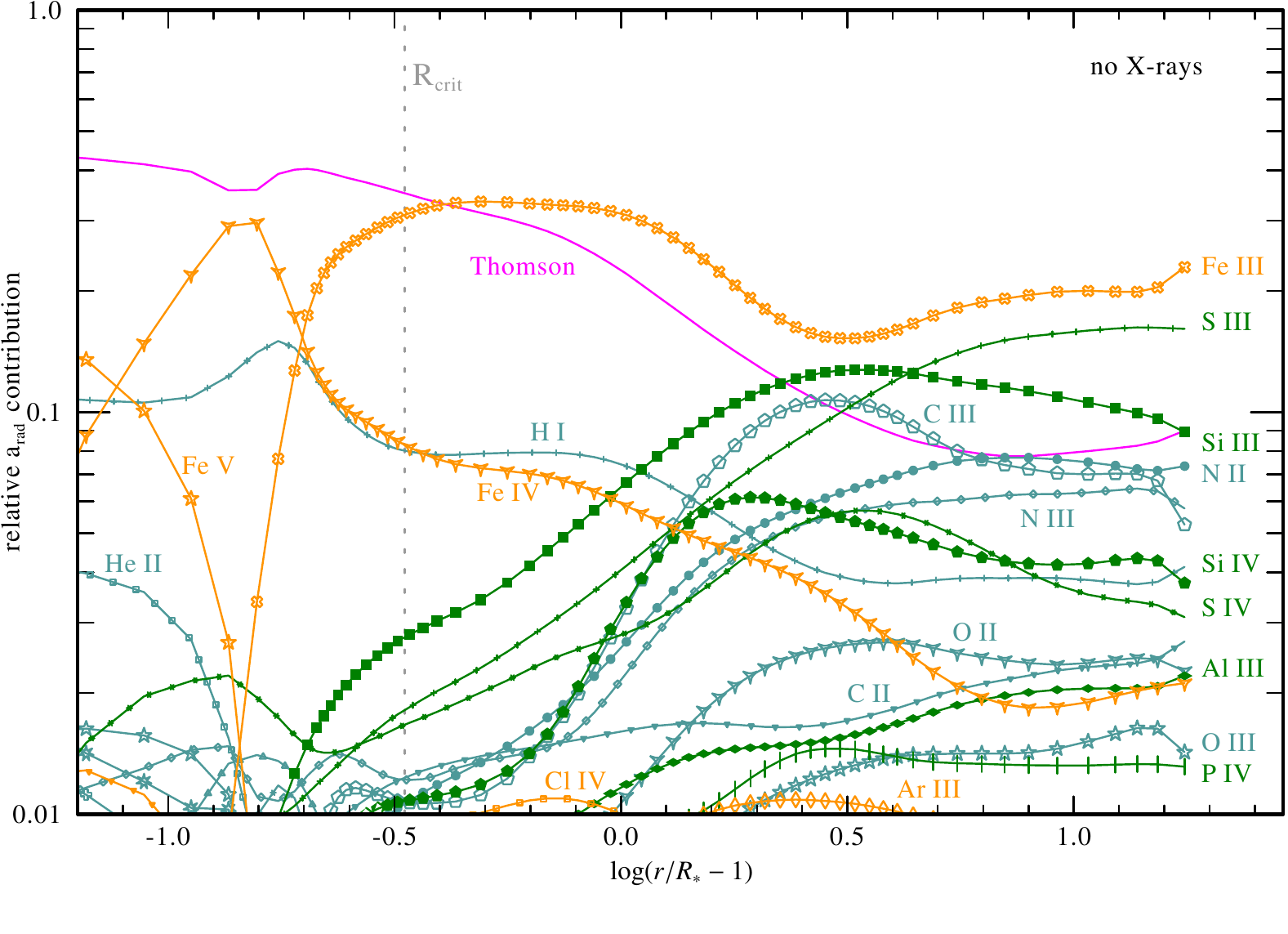}}
  \caption{Same as Fig.\,\ref{fig:leadions}, but for the HD consistent model without any X-ray illumination.}
  \label{fig:leadionsnox}
\end{figure*}

        A more detailed look into the wind driving is possible through
        studying the contributions of the various elements to the radiative acceleration, as illustrated in Fig.\,\ref{fig:eclipseelem}
        for the model using the moderate X-ray flux. At solar metallicity, the iron opacities are
        the main line contributions to the driving until the sonic point and farther outward up to about $2\,R_\ast$. While
        this appears to be rather similar to the much hotter O-supergiant ($\zeta$\,Pup) model from \citet{Sander+2017}, it is noteworthy that
        the iron contribution already outweighs the free electron (Thomson) scattering contribution at the sonic point in the case of
        Vela X-1. This is not the case for $\zeta\,$Pup, where the main ion contributing to the iron acceleration fraction 
        is \ion{Fe}{v} instead of \ion{Fe}{iii} here, which is shown in Fig.\,\ref{fig:leadions}, where the relative contributions
        of the various ions in the wind are plotted. The different temperature and thus ionization regime also leads to the
         differences in the list of elements that provide major contributions in the outer wind: After a few
        $R_\ast$, sulfur, nitrogen, and the iron group (in descending order) all provide between $15\%$ and $20\%$ of the radiative force.
        Slightly smaller contributions stem from Si and C. Oxygen provides about $5\%,$ which is already below the Thomson contribution.
        Interestingly, it is followed by hydrogen -- because of its significant continuum contribution -- and then several elements (Al, Cl, P, and Ar) that contribute
        on a percentage level. The driving influence of K is already two orders of magnitude lower than that of the leading elements
        and below the basically negligible gas pressure. The remaining elements (He, Ne, Mg, and Ca) contribute even less and
        thus can be neglected with regard to driving. However, as the comparison with the $\zeta\,$Pup model from \citet{Sander+2017} illustrates, 
        we have to be cautious about generalizing these results. The influence of the various elements strongly depends 
        on the abundances and the ionization stages, and thus the picture can change drastically when transitioning to other temperature
        and/or abundance regimes. 
        
        As previously mentioned in Sect.\,\ref{sec:mdot}, our self-consistent model provides an interesting test-case for the
        discussion of the (hot) bistability jump, which was originally discovered in wind models for P\,Cyg by \citet{PP1990} and was
				later observationally corroborated by \citet{Lamers+1995}, who found a discontinuity in the $\varv_\infty/\varv_\text{esc}$ ratio 
				around an effective temperature of $21\,$kK.  \citet{PP1990} have attributed the nature of the jump to a change in the leading ionization, which can
        have a significant effect on the driving. \citet{Vink+1999} were able to find the bistability jump in their theoretical models by obtaining the radiative
        acceleration through Monte Carlo calculations, but at the slightly higher temperature of $25\,$kK. Using grids of CMFGEN models and interpolating at 
        $Q = 1$, \citet{Petrov+2016} predicted the region of the jump to be located between $22.5\,$kK and $20\,$kK. In both approaches, the change
        in the ionization structure is identified as the origin of the jump. Detailed knowledge about the exact locations and the intensity
        of the bistability jumps is very important, not only for our understanding of key objects like B-hypergiants
        \citep[e.g.,][]{Clark+2012,Oskinova+2017} but also for correctly modeling the evolution of massive OB stars in general, for example, as recently demonstrated
        by \citet{Keszthelyi+2017}.
        
        With our result of $T_{2/3} = 23.5\,$kK, we are still very close to the bistability jump, which is also indicated by the unusual combination 
        of a low $\dot{M}$ combined with $\varv_\infty/\varv_\text{esc} \approx 1.3$ (see Sect.\,\ref{sec:strat}). In the model without X-rays, \ion{Fe}{iv} is the leading
        Fe ion with regard to the population numbers, while Fig.\,\ref{fig:leadionsnox} clearly shows that its contribution to the driving in the wind 
        is clearly minor compared to \ion{Fe}{iii}, which was also found in calculations from both \citet{Vink+1999} and \citet{Petrov+2016}. 
        While the results visualized in Figs.\,\ref{fig:eclipseelem} and \ref{fig:leadions} are for a moderate X-ray illumination, the results for the
        non-X-ray model differ only in the details, as can be seen when comparing Fig.\,\ref{fig:leadions} with Fig.\,\ref{fig:leadionsnox}. The absence of all 
        the higher ions (\ion{C}{iv}, \ion{N}{iv}, \ion{N}{v}, also \ion{O}{iv} and \ion{Cl}{iv}  at greater distance) in the wind part is conspicuous,
        since these ions are not significantly populated without the X-ray channel. In absolute numbers, the picture is very similar to Fig.\,\ref{fig:eclipseelem},
        but without X-rays, the contributions of C, N, O, Si, and S are smaller, and Si and S are only mildly affected. Nevertheless, this change is enough
        to barely make iron - or the iron group elements, to be more exact - the leading element now. 

        From the calculations of \citet{Petrov+2016}, the contribution of CNO elements to the total work ratio was also derived. With our detailed calculations
        and their visualized results in Figs.\,\ref{fig:eclipseelem} and \ref{fig:leadionsnox}, we can now not only verify their finding that the combined 
        CNO contribution outweighs the iron-driving contribution in our temperature regime, but investigate this also in a depth-dependent manner,
        revealing that this is is only true for the outer wind regime at $r \gtrsim 3\,R_\ast$, while \ion{Fe}{iii} and electron scattering are the dominant
        drivers below, down to the regime around the sonic point. We also discovered the important influence of Si and S in accelerating the wind, in
        particular \ion{S}{iii} and \ion{Si}{iii}, but also \ion{S}{iv} and \ion{Si}{iv} to a lesser degree, which was not noted by \citet{Petrov+2016} 
				although they included these elements in their calculations. Given these results and the
        significant differences between the contributions of C, N, and O, we therefore conclude that it is dangerous to consider the CNO elements alone
        as important drivers and disregard the importance of other elements such as S and Si. While S and N have comparable influences, the influence
        of O is about a factor of three lower. Of course we have to keep in mind that nitrogen is enriched while oxygen is slightly depleted, which is likely
        a result of the CNO cycle, but carbon is depleted as well, even much more strongly than oxygen, and still it has about twice the 
        driving influence of oxygen in the wind. These deviations from the solar composition might be another reason why our results differ from predictions
        such as those in \citet{Vink+2000}. We plan to investigate the whole topic of detailed driving contributions in more detail in future studies.
        
        We finally consider the origin of the ``bump'' structure that occurs in the radiative acceleration regardless of X-ray 
        illumination shortly below the sonic point. This structure, which is most notably seen in the contributions from iron, hydrogen, and electron scattering,
        does not originate from any major ionization change or non-monotonic feature in the temperature structure, but instead stems from the increase in the
        clumping factor in this region. Below the bump, we have $D \approx 1$, while we have $D \approx 7$ at the critical point. While the further increase
        outward does not leave a notable substructure on the radiative acceleration, the onset of the clumping does, in contrast to our O supergiant model
        in \citet{Sander+2017}, even though the clumping increase there started below the critical point as well. The influence of the clumping stratification
        on the radiative acceleration is another topic that has to be studied in more detail in the future.

\section{Conclusions}
  \label{sec:conclusions}
        
        We constructed HD consistent PoWR models for the donor star of Vela X-1, HD 77581, thereby
        for the first time applying our recently introduced next-generation PoWR models \citep{Sander+2017} to the
        regime of early-B supergiants. The values of the stellar parameters were motivated by the previous empirical study from \citet{GimenezGarcia+2016},
        and the resulting models reproduce the overall spectral appearance of the donor star. Three HD 
        consistent models using different levels of X-ray illuminations demonstrate the effects of
        the X-rays that arise from accretion onto the NS in the Vela X-1 system on the donor star wind. 
        
        Our atmosphere models prove that the low terminal velocity derived by \citet{GimenezGarcia+2016}
        is consistent with the radiative acceleration provided by the radiation of the donor star, in
        line with earlier predictions by \citet{Krticka+2012}. In the inner wind region, our hydrodynamical models yield a stratification
        that is notably different from what is obtained when a prescribed $\beta$-law is used. Our 
        calculations furthermore reveal that a certain amount of X-rays influences the ionization balance such that additional
        driving is provided in the outer wind, and the terminal velocity is increased by about 10\% compared to a similar donor star
        without X-rays. However, when the X-ray illumination is too high, a breakdown of the acceleration occurs in the outer wind.
        It is noteworthy that this breakdown does not occur already at the distance of the NS $d_\textsc{ns}$, but instead 
        much farther out after $r \gtrsim 3\,R_\ast \approx 2 d_\textsc{ns}$. Nevertheless, owing to the restriction to a stationary
        1D description of the wind and a rather approximate X-ray treatment, this result should be taken with care.
        
        Our calculations confirm the empirically derived mass-loss rate of the donor star of Vela X-1 of $\log \dot{M} \approx -6.2$
        assuming a depth-dependent microclumping with $D_\infty = 11$. The X-ray illumination has only very little influence on the wind
        mass loss, potentially increasing the rate by up to $0.1\,$dex in the direction toward the NS. The wind velocity
        in the inner wind and especially at the distance of the NS $\varv(d_\textsc{ns}) \approx 100\,$km\,s$^{-1}$ is lower than typically 
        estimated from prescribed $\beta$-laws. Our obtained $\varv(d_\textsc{ns})$ is lower than the orbital speed of the NS, but an estimate
        assuming direct Bondi-Hoyle-Littleton accretion yields excellent agreement between the mean observed X-ray luminosity of Vela X-1
        and our prediction. Tables with the stratifications from all the HD consistent models are
        provided in Appendix \ref{sec:apptables}.
        
        A detailed inspection of the driving contributions reveals that a plethora of ions from more than $\text{ten}$ different elements need to be 
        considered to properly reconstruct the full radiative wind acceleration. The leading ion in our early B-type supergiant 
        wind is \ion{Fe}{iii}, which contributes about 15\% in the case of a moderate X-ray illumination. In the
        outer wind, \ion{S}{iii} reaches an almost comparable fraction, followed by \ion{Si}{iii} and \ion{C}{iii} which contribute 
        about 10\% in the wind. Although the general picture of the B-supergiant wind shows similarities to
        our previous O-supergiant results \citep{Sander+2017}, the detailed contributions are significantly different because of the different
        stellar parameters. Further studies are required before more general conclusions should be drawn.

\begin{acknowledgements}
  We would like to thank the anonymous referee for the insightful questions and suggestions
	that helped to strengthen this paper. A.A.C. Sander is supported by
  the Deutsche Forschungsgemeinschaft (DFG) under grant HA 1455/26. T.S. acknowledges support from the 
  German ``Verbundforschung'' (DLR) grant 50 OR 1612.
  This research made use of the SIMBAD and VizieR databases, operated at CDS, Strasbourg, France. 
  This publication was motivated by a team meeting sponsored
  by the International Space Science Institute at Bern, Switzerland.
  We acknowledge support from the Faculty of the European Space Astronomy Centre (ESAC).
\end{acknowledgements}


\bibliographystyle{aa} 
\bibliography{hydrovelax1}
  
\begin{appendix}

\onecolumn

\section{List of atomic data used in the stellar atmosphere models}
  \label{sec:appdatom}
                                        
\begin{table*}[h!]
  \caption{Atomic data used in the HD models. The numbers in parentheses refer to the model with the high X-ray illumination,
                 which requires the inclusion of higher ionization stages.}
  \label{tab:datom}
  \centering
  \begin{tabular}{l p{1mm} c c p{1mm} c c p{5mm} l p{1mm} c c p{1mm} c c}
  \hline\hline
    Ion  \rule[0mm]{0mm}{3mm}               & &  \multicolumn{2}{c}{Levels} & &  \multicolumn{2}{c}{Lines\tablefootmark{a}}   &    &
    Ion  \rule[0mm]{0mm}{3mm}               & &  \multicolumn{2}{c}{Levels} & &  \multicolumn{2}{c}{Lines\tablefootmark{a}}  \\
  \hline

    \ion{H}{i}  \rule[0mm]{0mm}{3mm} &  & \multicolumn{2}{c}{22}  &  &  \multicolumn{2}{c}{231}   &  &    \ion{Si}{iii}  &  & \multicolumn{2}{c}{24}   &  & \multicolumn{2}{c}{ 69}  \\
    \ion{H}{ii}                      &  & \multicolumn{2}{c}{ 1}  &  &  \multicolumn{2}{c}{  0}   &  &    \ion{Si}{iv}   &  &       23   &      (23)   &  &       72     &     (69)  \\
    \ion{He}{i}                      &  & \multicolumn{2}{c}{35}  &  &  \multicolumn{2}{c}{271}   &  &    \ion{Si}{v}    &  &       11   &      (23)   &  &       11     &     (48)  \\
    \ion{He}{ii}                     &  & \multicolumn{2}{c}{26}  &  &  \multicolumn{2}{c}{325}   &  &    \ion{Si}{vi}   &  &        0   &       (1)   &  &        0     &      (0)  \\
    \ion{He}{iii}                    &  & \multicolumn{2}{c}{ 1}  &  &  \multicolumn{2}{c}{  0}   &  &    \ion{P}{iii}   &  &        1   &      (47)   &  &        0     &    (175)  \\   
    \ion{C}{i}                       &  &       15   &      (10)  &  &          30   &     (10)   &  &    \ion{P}{iv}    &  & \multicolumn{2}{c}{12}   &  & \multicolumn{2}{c}{ 16}  \\   
    \ion{C}{ii}                      &  & \multicolumn{2}{c}{32}  &  &  \multicolumn{2}{c}{148}   &  &    \ion{P}{v}     &  & \multicolumn{2}{c}{11}   &  & \multicolumn{2}{c}{ 22}  \\   
    \ion{C}{iii}                     &  & \multicolumn{2}{c}{40}  &  &  \multicolumn{2}{c}{226}   &  &    \ion{P}{vi}    &  & \multicolumn{2}{c}{ 1}   &  & \multicolumn{2}{c}{  0}  \\   
    \ion{C}{iv}                      &  & \multicolumn{2}{c}{25}  &  &  \multicolumn{2}{c}{230}   &  &    \ion{S}{i}     &  &       30   &       (5)   &  &       76     &      (4)  \\   
    \ion{C}{v}                       &  & \multicolumn{2}{c}{10}  &  &  \multicolumn{2}{c}{ 13}   &  &    \ion{S}{ii}    &  &       32   &      (15)   &  &       80     &     (18)  \\
    \ion{C}{vi}                      &  &        1   &      (10)  &  &          10   &     (45)   &  &    \ion{S}{iii}   &  & \multicolumn{2}{c}{23}   &  & \multicolumn{2}{c}{ 38}  \\   
    \ion{C}{vii}                     &  &        0   &       (1)  &  &           0   &      (0)   &  &    \ion{S}{iv}    &  & \multicolumn{2}{c}{25}   &  & \multicolumn{2}{c}{ 54}  \\
    \ion{N}{i}                       &  & \multicolumn{2}{c}{10}  &  &  \multicolumn{2}{c}{ 13}   &  &    \ion{S}{v}     &  & \multicolumn{2}{c}{10}   &  & \multicolumn{2}{c}{ 13}  \\
    \ion{N}{ii}                      &  &       38   &      (20)  &  &         201   &     (29)   &  &    \ion{S}{vi}    &  &       22   &      (10)   &  &       75     &     (21)  \\   
    \ion{N}{iii}                     &  & \multicolumn{2}{c}{30}  &  &  \multicolumn{2}{c}{ 94}   &  &    \ion{S}{vii}   &  &        0   &       (1)   &  &        0     &      (0)  \\
    \ion{N}{iv}                      &  & \multicolumn{2}{c}{38}  &  &  \multicolumn{2}{c}{154}   &  &    \ion{Cl}{iii}  &  & \multicolumn{2}{c}{ 1}   &  & \multicolumn{2}{c}{  0}  \\
    \ion{N}{v}                       &  & \multicolumn{2}{c}{20}  &  &  \multicolumn{2}{c}{114}   &  &    \ion{Cl}{vi}   &  &       24   &      (15)   &  &       34     &     (17)  \\
    \ion{N}{vi}                      &  & \multicolumn{2}{c}{14}  &  &  \multicolumn{2}{c}{ 48}   &  &    \ion{Cl}{v}    &  & \multicolumn{2}{c}{18}   &  & \multicolumn{2}{c}{ 29}  \\
    \ion{N}{vii}                     &  &        1   &      (10)  &  &           0   &      (7)   &  &    \ion{Cl}{vi}   &  & \multicolumn{2}{c}{23}   &  & \multicolumn{2}{c}{ 46}  \\ 
    \ion{N}{viii}                    &  &        0   &       (1)  &  &           0   &      (0)   &  &    \ion{Cl}{vii}  &  & \multicolumn{2}{c}{ 1}   &  & \multicolumn{2}{c}{  0}  \\        
    \ion{O}{i}                       &  &       13   &      (10)  &  &          15   &      (9)   &  &    \ion{Ar}{i}    &  &       14   &      (10)   &  &       34     &     (17)  \\   
    \ion{O}{ii}                      &  & \multicolumn{2}{c}{37}  &  &  \multicolumn{2}{c}{150}   &  &    \ion{Ar}{ii}   &  &       20   &      (10)   &  &       33     &      (9)  \\
    \ion{O}{iii}                     &  & \multicolumn{2}{c}{33}  &  &  \multicolumn{2}{c}{121}   &  &    \ion{Ar}{iii}  &  &       14   &      (10)   &  &       13     &      (8)  \\
    \ion{O}{iv}                      &  & \multicolumn{2}{c}{29}  &  &  \multicolumn{2}{c}{ 77}   &  &    \ion{Ar}{iv}   &  & \multicolumn{2}{c}{13}   &  & \multicolumn{2}{c}{ 20}  \\
    \ion{O}{v}                       &  &       36   &      (54)  &  &         153   &    (260)   &  &    \ion{Ar}{v}    &  & \multicolumn{2}{c}{10}   &  & \multicolumn{2}{c}{ 11}  \\
    \ion{O}{vi}                      &  & \multicolumn{2}{c}{16}  &  &  \multicolumn{2}{c}{101}   &  &    \ion{Ar}{vi}   &  & \multicolumn{2}{c}{ 9}   &  & \multicolumn{2}{c}{ 11}  \\
    \ion{O}{vii}                     &  &        1   &      (15)  &  &           0   &     (64)   &  &    \ion{Ar}{vii}  &  & \multicolumn{2}{c}{20}   &  & \multicolumn{2}{c}{ 34}  \\  
    \ion{O}{viii}                    &  &        0   &      (15)  &  &           0   &    (105)   &  &    \ion{Ar}{viii} &  & \multicolumn{2}{c}{ 1}   &  & \multicolumn{2}{c}{  0}  \\   
    \ion{O}{ix}                      &  &        0   &       (1)  &  &           0   &      (0)   &  &    \ion{K}{i}     &  &       20   &      (15)   &  &       48     &     (32)  \\ 
    \ion{Ne}{i}                      &  &        8   &      (10)  &  &          14   &     (26)   &  &    \ion{K}{ii}    &  &       20   &      (15)   &  &       56     &     (30)  \\  
    \ion{Ne}{ii}                     &  &       18   &      (10)  &  &          40   &      (9)   &  &    \ion{K}{iii}   &  &       20   &      (10)   &  &       40     &     (12)  \\ 
    \ion{Ne}{iii}                    &  & \multicolumn{2}{c}{18}  &  &  \multicolumn{2}{c}{ 18}   &  &    \ion{K}{iv}    &  &       23   &      (10)   &  &       27     &      (9)  \\   
    \ion{Ne}{iv}                     &  &       35   &      (20)  &  &         159   &     (26)   &  &    \ion{K}{v}     &  &       19   &      (10)   &  &       33     &     (16)  \\ 
    \ion{Ne}{v}                      &  & \multicolumn{2}{c}{20}  &  &  \multicolumn{2}{c}{ 23}   &  &    \ion{K}{vi}    &  & \multicolumn{2}{c}{ 1}   &  & \multicolumn{2}{c}{  0}  \\
    \ion{Ne}{vi}                     &  & \multicolumn{2}{c}{20}  &  &  \multicolumn{2}{c}{ 35}   &  &    \ion{Ca}{i}    &  &       20   &      (15)   &  &       35     &     (24)  \\ 
    \ion{Ne}{vii}                    &  &        1   &      (10)  &  &           0   &     (11)   &  &    \ion{Ca}{ii}   &  &       20   &      (15)   &  &       48     &     (31)  \\ 
    \ion{Ne}{viii}                   &  &        0   &      (10)  &  &           0   &     (20)   &  &    \ion{Ca}{iii}  &  & \multicolumn{2}{c}{14}   &  & \multicolumn{2}{c}{ 18}  \\  
    \ion{Ne}{ix}                     &  &        0   &      (10)  &  &           0   &     (13)   &  &    \ion{Ca}{iv}   &  & \multicolumn{2}{c}{24}   &  & \multicolumn{2}{c}{ 43}  \\
    \ion{Ne}{x}                      &  &        0   &      (10)  &  &           0   &     (13)   &  &    \ion{Ca}{v}    &  & \multicolumn{2}{c}{15}   &  & \multicolumn{2}{c}{ 12}  \\  
    \ion{Ne}{xi}                     &  &        0   &       (1)  &  &           0   &      (0)   &  &    \ion{Ca}{vi}   &  & \multicolumn{2}{c}{15}   &  & \multicolumn{2}{c}{ 17}  \\ 
    \ion{Mg}{i}                      &  & \multicolumn{2}{c}{ 1}  &  &  \multicolumn{2}{c}{  0}   &  &    \ion{Ca}{vii}  &  & \multicolumn{2}{c}{20}   &  & \multicolumn{2}{c}{ 28}  \\  
    \ion{Mg}{ii}                     &  & \multicolumn{2}{c}{20}  &  &  \multicolumn{2}{c}{ 57}   &  &    \ion{Ca}{viii} &  & \multicolumn{2}{c}{ 1}   &  & \multicolumn{2}{c}{  0}  \\  
    \ion{Mg}{iii}                    &  & \multicolumn{2}{c}{20}  &  &  \multicolumn{2}{c}{ 41}   &  &    \ion{Fe\tablefootmark{b}}{i}    & & \multicolumn{2}{c}{ 1} & & \multicolumn{2}{c}{  0}  \\  
    \ion{Mg}{iv}                     &  & \multicolumn{2}{c}{17}  &  &  \multicolumn{2}{c}{ 27}   &  &    \ion{Fe\tablefootmark{b}}{ii}   & & \multicolumn{2}{c}{~3 {\small [15\,077]}} & & \multicolumn{2}{c}{~~2 {\small [3\,688\,311]}}   \\ 
    \ion{Al}{i}                      &  &       10   &       (1)  &  &          16   &      (0)   &  &    \ion{Fe\tablefootmark{b}}{iii}  & & \multicolumn{2}{c}{13 {\small [15\,233]}} & & \multicolumn{2}{c}{~40 {\small [2\,888\,649]}}   \\
    \ion{Al}{ii}                     &  & \multicolumn{2}{c}{10}  &  &  \multicolumn{2}{c}{ 11}   &  &    \ion{Fe\tablefootmark{b}}{iv}   & & \multicolumn{2}{c}{18 {\small [30\,302]}} & & \multicolumn{2}{c}{~77 {\small [7\,230\,222]}}   \\
    \ion{Al}{iii}                    &  & \multicolumn{2}{c}{10}  &  &  \multicolumn{2}{c}{ 18}   &  &    \ion{Fe\tablefootmark{b}}{v}    & & \multicolumn{2}{c}{22 {\small [20\,639]}} & & \multicolumn{2}{c}{107 {\small [5\,924\,318]}}   \\
    \ion{Al}{iv}                     &  & \multicolumn{2}{c}{10}  &  &  \multicolumn{2}{c}{ 10}   &  &    \ion{Fe\tablefootmark{b}}{vi}   & & \multicolumn{2}{c}{29 {\small [17\,728]}} & & \multicolumn{2}{c}{194 {\small [3\,948\,149]}}   \\
    \ion{Al}{v}                      &  & \multicolumn{2}{c}{10}  &  &  \multicolumn{2}{c}{  9}   &  &    \ion{Fe\tablefootmark{b}}{vii}  & & \multicolumn{2}{c}{19 {\small [16\,752]}} & & \multicolumn{2}{c}{~87 {\small [4\,002\,080]}}   \\ 
    \ion{Al}{vi}                     &  & \multicolumn{2}{c}{ 1}  &  &  \multicolumn{2}{c}{  0}   &  &    \ion{Fe\tablefootmark{b}}{viii} & & \multicolumn{2}{c}{ 1} & & \multicolumn{2}{c}{  0}   \\ 
    \ion{Si}{i}                      &  &       20   &       (3)  &  &          45   &      (2)   &  &                   &  &          &             & &             &           \\
    \ion{Si}{ii}                     &  & \multicolumn{2}{c}{20}  &  &  \multicolumn{2}{c}{ 35}   &  &       Total       &  &    1524  &    (1486)   & &     4976    &   (4783)  \\    
  \hline
  \end{tabular}
  \tablefoot{
        \tablefoottext{a}{Number of transitions with non-negligible oscillator strengths as considered in the radiative transfer calculations}
        \tablefoottext{b}{For Fe, the large numbers refer to superlevels and superline transitions that are used to cope with the enormous number of lines and transitions given as smaller numbers in brackets. The atom listed as Fe here is a generic element that also includes the iron group elements Sc, Ti, V, Cr, Mn, Co, and Ni \citep[see][for details]{GKH2002}.}
  }  
\end{table*}

\section{Stratifications of our hydrodynamcially consistent wind models for the donor star of Vela X-1}
  \label{sec:apptables}
        
\begin{longtable}{c c c c c}
  \caption{\label{apptab:nolx} Stratification for the atmosphere model without any X-ray illumination}\\
    \hline\hline
  $r [R_\ast]\,-1$ \rule[0mm]{0mm}{3.5mm} & $\varv$  [km\,s$^{-1}$] & $T_\mathrm{e}$ [kK] & $\log\,(n_\mathrm{tot}$ [cm$^{-3}$]$)$ & $\log\,(n_\mathrm{e}$ [cm$^{-3}$]$)$  \\
    \hline
        \endfirsthead
    \caption{continued.}\\
    \hline\hline
  $r [R_\ast]\,-1$ \rule[0mm]{0mm}{3.5mm} & $\varv$  [km\,s$^{-1}$] & $T_\mathrm{e}$ [kK] & $\log\,(n_\mathrm{tot}$ [cm$^{-3}$]$)$ & $\log\,(n_\mathrm{e}$ [cm$^{-3}$]$)$  \\
\hline
\endhead
\hline
\endfoot
     19.0     &  531.7     & 11.209 &    7.239      &     7.240     \\
     16.2     &  525.3     & 12.449 &    7.378      &     7.378     \\
     14.5     &  520.2     & 12.620 &    7.471      &     7.472     \\
     13.1     &  514.9     & 12.421 &    7.555      &     7.555     \\
     11.6     &  507.3     & 12.116 &    7.662      &     7.662     \\
     9.92     &  496.2     & 11.790 &    7.795      &     7.795     \\
     8.46     &  482.7     & 11.666 &    7.932      &     7.932     \\
     7.44     &  469.9     & 11.798 &    8.042      &     8.043     \\
     6.64     &  457.3     & 12.052 &    8.141      &     8.141     \\
     5.88     &  442.3     & 12.447 &    8.246      &     8.246     \\
     5.18     &  424.9     & 13.074 &    8.357      &     8.358     \\
     4.61     &  407.7     & 13.917 &    8.459      &     8.459     \\
     4.23     &  393.9     & 14.654 &    8.535      &     8.536     \\
     3.93     &  381.9     & 15.268 &    8.600      &     8.600     \\
     3.66     &  369.9     & 15.734 &    8.661      &     8.662     \\
     3.41     &  357.4     & 16.244 &    8.724      &     8.725     \\
     3.16     &  343.5     & 16.643 &    8.792      &     8.793     \\
     2.93     &  329.0     & 16.992 &    8.862      &     8.862     \\
     2.72     &  315.0     & 17.268 &    8.927      &     8.928     \\
     2.53     &  300.6     & 17.439 &    8.994      &     8.994     \\
     2.34     &  284.9     & 17.506 &    9.066      &     9.066     \\
     2.16     &  268.8     & 17.466 &    9.139      &     9.139     \\
     2.00     &  253.7     & 17.371 &    9.208      &     9.208     \\
     1.86     &  238.6     & 17.221 &    9.277      &     9.278     \\
     1.71     &  222.3     & 17.025 &    9.353      &     9.353     \\
     1.58     &  206.1     & 16.832 &    9.430      &     9.430     \\
     1.47     &  191.3     & 16.703 &    9.502      &     9.502     \\
     1.36     &  176.9     & 16.653 &    9.574      &     9.575     \\
     1.25     &  161.8     & 16.658 &    9.653      &     9.653     \\
     1.15     &  147.3     & 16.716 &    9.733      &     9.733     \\
     1.07     &  134.4     & 16.843 &    9.808      &     9.808     \\
    0.988     &  122.3     & 17.033 &    9.883      &     9.883     \\
    0.909     &  110.1     & 17.258 &    9.964      &     9.964     \\
    0.836     &  98.62     & 17.490 &   10.045      &    10.046     \\
    0.774     &  88.84     & 17.718 &   10.121      &    10.121     \\
    0.716     &  79.90     & 17.926 &   10.195      &    10.196     \\
    0.659     &  70.97     & 18.111 &   10.276      &    10.277     \\
    0.597     &  61.17     & 18.269 &   10.374      &    10.374     \\
    0.525     &  49.51     & 18.400 &   10.506      &    10.506     \\
    0.456     &  37.76     & 18.527 &   10.664      &    10.664     \\
    0.407     &  29.12     & 18.686 &   10.806      &    10.807     \\
    0.378     &  23.71     & 18.848 &   10.914      &    10.914     \\
    0.353     &  19.18     & 19.022 &   11.022      &    11.022     \\
    0.330     &  15.00     & 19.217 &   11.143      &    11.144     \\
    0.313     &  12.03     & 19.390 &   11.251      &    11.251     \\
    0.298     &  9.653     & 19.553 &   11.356      &    11.356     \\
    0.282     &  7.249     & 19.736 &   11.491      &    11.492     \\
    0.266     &  5.236     & 19.912 &   11.643      &    11.644     \\
    0.253     &  3.842     & 20.059 &   11.787      &    11.787     \\
    0.242     &  2.869     & 20.207 &   11.921      &    11.921     \\
    0.233     &  2.172     & 20.397 &   12.049      &    12.049     \\
    0.225     &  1.662     & 20.631 &   12.170      &    12.171     \\
    0.217     &  1.282     & 20.940 &   12.289      &    12.289     \\
    0.209     & 0.9318     & 21.409 &   12.433      &    12.434     \\
    0.197     & 0.5911     & 22.090 &   12.639      &    12.640     \\
    0.183     & 0.3463     & 23.810 &   12.882      &    12.882     \\
    0.167     & 0.2219     & 26.160 &   13.087      &    13.088     \\
    0.147     & 0.1680     & 27.980 &   13.222      &    13.228     \\
    0.125     & 0.1492     & 31.635 &   13.291      &    13.310     \\
    0.100     & 0.1173     & 35.394 &   13.415      &    13.449     \\
    0.767E-01 & 0.8363E-01 & 39.214 &   13.581      &    13.624     \\
    0.551E-01 & 0.5715E-01 & 42.973 &   13.764      &    13.810     \\
    0.354E-01 & 0.3785E-01 & 46.686 &   13.959      &    14.007     \\
    0.194E-01 & 0.2584E-01 & 50.354 &   14.138      &    14.186     \\
    0.954E-02 & 0.2002E-01 & 53.118 &   14.257      &    14.305     \\
    0.424E-02 & 0.1743E-01 & 54.807 &   14.322      &    14.370     \\
    0.212E-02 & 0.1649E-01 & 55.523 &   14.348      &    14.396     \\
    0.106E-02 & 0.1599E-01 & 55.913 &   14.362      &    14.411     \\
    0.531E-03 & 0.1571E-01 & 55.970 &   14.370      &    14.419     \\
     0.00     & 0.1562E-01 & 56.148 &   14.373      &    14.422     \\
\end{longtable}

\begin{longtable}{c c c c c}
  \caption{\label{apptab:eclipselx} Stratification for the atmosphere model with moderate X-ray illumination ($L_\text{X} \approx 10^{33}\,$erg/s)}\\
    \hline\hline
  $r [R_\ast]\,-1$ \rule[0mm]{0mm}{3.5mm} & $\varv$  [km\,s$^{-1}$] & $T_\mathrm{e}$ [kK] & $\log\,(n_\mathrm{tot}$ [cm$^{-3}$]$)$ & $\log\,(n_\mathrm{e}$ [cm$^{-3}$]$)$  \\
    \hline
        \endfirsthead
    \caption{continued.}\\
    \hline\hline
  $r [R_\ast]\,-1$ \rule[0mm]{0mm}{3.5mm} & $\varv$  [km\,s$^{-1}$] & $T_\mathrm{e}$ [kK] & $\log\,(n_\mathrm{tot}$ [cm$^{-3}$]$)$ & $\log\,(n_\mathrm{e}$ [cm$^{-3}$]$)$  \\
\hline
\endhead
\hline
\endfoot
     19.0     &  584.5     & 11.209 &    7.212      &     7.212     \\
     16.5     &  578.3     & 12.192 &    7.330      &     7.331     \\
     14.4     &  571.1     & 12.576 &    7.447      &     7.448     \\
     12.8     &  563.4     & 12.402 &    7.553      &     7.553     \\
     11.2     &  553.9     & 12.107 &    7.667      &     7.667     \\
     9.48     &  540.2     & 11.803 &    7.807      &     7.808     \\
     8.03     &  523.7     & 11.704 &    7.950      &     7.951     \\
     7.02     &  508.3     & 11.835 &    8.066      &     8.066     \\
     6.25     &  493.0     & 12.086 &    8.168      &     8.168     \\
     5.61     &  477.7     & 12.429 &    8.261      &     8.261     \\
     5.18     &  465.2     & 12.775 &    8.331      &     8.332     \\
     4.84     &  454.2     & 13.147 &    8.391      &     8.391     \\
     4.53     &  442.9     & 13.594 &    8.449      &     8.449     \\
     4.24     &  430.9     & 14.112 &    8.508      &     8.509     \\
     3.94     &  417.5     & 14.707 &    8.572      &     8.573     \\
     3.67     &  403.3     & 15.275 &    8.637      &     8.638     \\
     3.42     &  389.3     & 15.786 &    8.700      &     8.701     \\
     3.18     &  374.7     & 16.242 &    8.764      &     8.764     \\
     2.95     &  358.5     & 16.652 &    8.833      &     8.833     \\
     2.73     &  341.7     & 17.003 &    8.903      &     8.904     \\
     2.54     &  325.6     & 17.264 &    8.970      &     8.970     \\
     2.36     &  309.2     & 17.439 &    9.037      &     9.037     \\
     2.19     &  291.5     & 17.485 &    9.110      &     9.110     \\
     2.02     &  273.5     & 17.452 &    9.184      &     9.184     \\
     1.88     &  256.9     & 17.357 &    9.253      &     9.253     \\
     1.75     &  240.6     & 17.208 &    9.322      &     9.323     \\
     1.61     &  223.2     & 17.020 &    9.397      &     9.398     \\
     1.49     &  206.2     & 16.838 &    9.473      &     9.474     \\
     1.39     &  191.0     & 16.706 &    9.544      &     9.545     \\
     1.29     &  176.5     & 16.661 &    9.614      &     9.615     \\
     1.19     &  161.6     & 16.666 &    9.690      &     9.690     \\
     1.10     &  147.4     & 16.718 &    9.766      &     9.767     \\
     1.03     &  135.0     & 16.833 &    9.837      &     9.837     \\
    0.957     &  123.5     & 17.001 &    9.906      &     9.907     \\
    0.886     &  112.0     & 17.208 &    9.980      &     9.981     \\
    0.807     &  98.98     & 17.477 &   10.071      &    10.072     \\
    0.710     &  83.22     & 17.834 &   10.194      &    10.195     \\
    0.613     &  67.31     & 18.151 &   10.337      &    10.338     \\
    0.530     &  53.40     & 18.346 &   10.484      &    10.484     \\
    0.460     &  41.09     & 18.489 &   10.638      &    10.639     \\
    0.409     &  31.61     & 18.632 &   10.783      &    10.784     \\
    0.377     &  25.42     & 18.792 &   10.897      &    10.898     \\
    0.350     &  20.20     & 18.968 &   11.014      &    11.015     \\
    0.324     &  15.31     & 19.179 &   11.152      &    11.152     \\
    0.305     &  11.83     & 19.379 &   11.276      &    11.277     \\
    0.289     &  9.120     & 19.569 &   11.400      &    11.401     \\
    0.271     &  6.478     & 19.779 &   11.561      &    11.562     \\
    0.254     &  4.382     & 19.981 &   11.742      &    11.743     \\
    0.240     &  3.034     & 20.169 &   11.912      &    11.912     \\
    0.229     &  2.158     & 20.393 &   12.067      &    12.068     \\
    0.220     &  1.569     & 20.693 &   12.213      &    12.213     \\
    0.212     &  1.167     & 21.069 &   12.347      &    12.348     \\
    0.204     & 0.8856     & 21.473 &   12.472      &    12.473     \\
    0.196     & 0.6579     & 22.015 &   12.607      &    12.608     \\
    0.185     & 0.4612     & 23.130 &   12.769      &    12.770     \\
    0.173     & 0.3227     & 24.656 &   12.933      &    12.934     \\
    0.160     & 0.2511     & 26.107 &   13.052      &    13.054     \\
    0.146     & 0.2198     & 27.468 &   13.121      &    13.126     \\
    0.128     & 0.1920     & 29.500 &   13.193      &    13.205     \\
    0.109     & 0.1482     & 32.389 &   13.320      &    13.342     \\
    0.886E-01 & 0.1047     & 35.787 &   13.487      &    13.520     \\
    0.690E-01 & 0.7372E-01 & 39.422 &   13.655      &    13.697     \\
    0.501E-01 & 0.5096E-01 & 43.149 &   13.831      &    13.877     \\
    0.324E-01 & 0.3425E-01 & 46.917 &   14.018      &    14.066     \\
    0.177E-01 & 0.2419E-01 & 50.535 &   14.182      &    14.230     \\
    0.777E-02 & 0.1890E-01 & 53.490 &   14.298      &    14.346     \\
    0.389E-02 & 0.1712E-01 & 54.771 &   14.344      &    14.392     \\
    0.194E-02 & 0.1629E-01 & 55.445 &   14.367      &    14.415     \\
    0.972E-03 & 0.1589E-01 & 55.797 &   14.379      &    14.427     \\
     0.00     & 0.1553E-01 & 56.141 &   14.390      &    14.438     \\
\end{longtable}

\begin{longtable}{c c c c c}
  \caption{\label{apptab:stronglx} Stratification for the atmosphere model with strong X-ray illumination ($L_\text{X} \approx 10^{37}\,$erg/s)}\\
    \hline\hline
  $r [R_\ast]\,-1$ \rule[0mm]{0mm}{3.5mm} & $\varv$  [km\,s$^{-1}$] & $T_\mathrm{e}$ [kK] & $\log\,(n_\mathrm{tot}$ [cm$^{-3}$]$)$ & $\log\,(n_\mathrm{e}$ [cm$^{-3}$]$)$  \\
    \hline
        \endfirsthead
    \caption{continued.}\\
    \hline\hline
  $r [R_\ast]\,-1$ \rule[0mm]{0mm}{3.5mm} & $\varv$  [km\,s$^{-1}$] & $T_\mathrm{e}$ [kK] & $\log\,(n_\mathrm{tot}$ [cm$^{-3}$]$)$ & $\log\,(n_\mathrm{e}$ [cm$^{-3}$]$)$  \\
\hline
\endhead
\hline
\endfoot
     19.0     &  377.6     & 11.209 &    7.504      &     7.553     \\
     16.6     &  377.6     & 12.177 &    7.614      &     7.663     \\
     15.2     &  377.6     & 12.410 &    7.688      &     7.737     \\
     14.0     &  377.6     & 12.336 &    7.756      &     7.805     \\
     12.5     &  377.6     & 11.864 &    7.843      &     7.892     \\
     10.9     &  377.6     & 11.800 &    7.953      &     8.002     \\
     9.22     &  377.6     & 11.916 &    8.087      &     8.136     \\
     7.77     &  377.6     & 12.705 &    8.220      &     8.269     \\
     6.77     &  377.6     & 13.650 &    8.325      &     8.374     \\
     6.00     &  377.6     & 14.637 &    8.416      &     8.464     \\
     5.28     &  377.6     & 15.596 &    8.510      &     8.558     \\
     4.69     &  377.4     & 16.278 &    8.595      &     8.643     \\
     4.31     &  376.1     & 16.688 &    8.657      &     8.704     \\
     4.02     &  373.4     & 16.969 &    8.709      &     8.753     \\
     3.74     &  368.1     & 17.205 &    8.766      &     8.802     \\
     3.46     &  361.2     & 17.437 &    8.826      &     8.841     \\
     3.22     &  354.2     & 17.452 &    8.883      &     8.887     \\
     3.00     &  346.8     & 17.452 &    8.940      &     8.943     \\
     2.77     &  338.1     & 17.450 &    9.002      &     9.004     \\
     2.56     &  328.4     & 17.430 &    9.065      &     9.067     \\
     2.37     &  318.3     & 17.327 &    9.125      &     9.127     \\
     2.20     &  307.2     & 17.119 &    9.185      &     9.188     \\
     2.03     &  294.3     & 16.938 &    9.252      &     9.254     \\
     1.87     &  280.3     & 16.805 &    9.320      &     9.323     \\
     1.73     &  266.6     & 16.682 &    9.384      &     9.387     \\
     1.61     &  252.7     & 16.675 &    9.449      &     9.451     \\
     1.48     &  237.2     & 16.676 &    9.519      &     9.521     \\
     1.36     &  221.5     & 16.680 &    9.591      &     9.593     \\
     1.26     &  207.0     & 16.705 &    9.658      &     9.660     \\
     1.17     &  192.7     & 16.847 &    9.725      &     9.727     \\
     1.08     &  177.6     & 17.074 &    9.798      &     9.800     \\
    0.975     &  159.6     & 17.349 &    9.889      &     9.891     \\
    0.866     &  139.4     & 17.622 &    9.997      &     9.999     \\
    0.781     &  122.2     & 17.830 &   10.095      &    10.096     \\
    0.717     &  108.8     & 17.980 &   10.177      &    10.178     \\
    0.664     &  97.46     & 18.104 &   10.252      &    10.252     \\
    0.617     &  87.30     & 18.206 &   10.325      &    10.325     \\
    0.569     &  77.12     & 18.314 &   10.405      &    10.405     \\
    0.518     &  66.13     & 18.417 &   10.500      &    10.501     \\
    0.469     &  55.26     & 18.525 &   10.607      &    10.607     \\
    0.431     &  46.85     & 18.629 &   10.701      &    10.702     \\
    0.399     &  39.58     & 18.721 &   10.794      &    10.795     \\
    0.364     &  31.44     & 18.888 &   10.916      &    10.917     \\
    0.330     &  23.80     & 19.125 &   11.059      &    11.060     \\
    0.303     &  17.92     & 19.357 &   11.200      &    11.201     \\
    0.281     &  13.50     & 19.558 &   11.338      &    11.338     \\
    0.262     &  10.17     & 19.725 &   11.473      &    11.474     \\
    0.247     &  7.684     & 19.891 &   11.606      &    11.607     \\
    0.233     &  5.815     & 20.056 &   11.737      &    11.737     \\
    0.218     &  4.113     & 20.250 &   11.898      &    11.899     \\
    0.197     &  2.410     & 20.768 &   12.145      &    12.145     \\
    0.175     &  1.203     & 21.984 &   12.463      &    12.464     \\
    0.151     & 0.5996     & 23.922 &   12.783      &    12.785     \\
    0.129     & 0.3945     & 25.915 &   12.981      &    12.985     \\
    0.114     & 0.3283     & 27.535 &   13.073      &    13.079     \\
    0.103     & 0.2813     & 28.934 &   13.149      &    13.159     \\
    0.924E-01 & 0.2300     & 30.455 &   13.245      &    13.259     \\
    0.805E-01 & 0.1753     & 32.534 &   13.372      &    13.392     \\
    0.649E-01 & 0.1220     & 35.618 &   13.542      &    13.571     \\
    0.505E-01 & 0.8628E-01 & 38.818 &   13.704      &    13.742     \\
    0.411E-01 & 0.6676E-01 & 41.043 &   13.824      &    13.866     \\
    0.343E-01 & 0.5398E-01 & 42.704 &   13.922      &    13.967     \\
    0.281E-01 & 0.4344E-01 & 44.349 &   14.021      &    14.067     \\
    0.210E-01 & 0.3245E-01 & 46.373 &   14.154      &    14.201     \\
    0.127E-01 & 0.2287E-01 & 49.132 &   14.313      &    14.361     \\
    0.573E-02 & 0.1785E-01 & 51.904 &   14.426      &    14.474     \\
    0.286E-02 & 0.1663E-01 & 53.177 &   14.460      &    14.508     \\
    0.143E-02 & 0.1615E-01 & 53.832 &   14.474      &    14.522     \\
    0.716E-03 & 0.1592E-01 & 54.161 &   14.481      &    14.529     \\
     0.00     & 0.1569E-01 & 54.483 &   14.487      &    14.535     \\
\end{longtable}

\end{appendix}

\end{document}